\documentclass[
aip,
reprint,
superscriptaddress,
amsmath,
amssymb,
]{revtex4-2}

\usepackage{graphicx}
\usepackage{epstopdf, epsfig}
\usepackage{bm}
\usepackage{amsmath}
\usepackage{amssymb}
\usepackage{subfigure}
\usepackage{color}
\usepackage{verbatim}
\usepackage{float}
\usepackage[dvipsnames]{xcolor}
\usepackage{natbib}

\newcommand{\M}{\mathcal{M}}
\newcommand{\bgrad}{\bm{\nabla}} %
\newcommand{\PP}{\mathcal{P}}

\begin{document}
	
	\title[Compressible LBM]{Extended Lattice Boltzmann Model for Gas Dynamics}
	
	\author{M. H. Saadat}
	\author{S. A. Hosseini}
	\author{B. Dorschner}
	\author{I. V. Karlin}\thanks{Corresponding author}
	\email{ikarlin@ethz.ch}
	\affiliation{Department of Mechanical and Process Engineering, ETH Zurich, 8092 Zurich, Switzerland}
	\date{\today}

\begin{abstract}
We propose a two-population lattice Boltzmann model on standard lattices for the simulation of compressible flows. The model is fully on-lattice and uses the single relaxation time Bhatnagar–Gross–Krook  kinetic equations along with appropriate correction terms to recover the Navier-Stokes-Fourier equations. The accuracy and performance of the model are analyzed through simulations of compressible benchmark cases including Sod shock tube, sound generation in shock-vortex interaction and compressible decaying turbulence in a box with eddy shocklets. It is demonstrated that the present model provides an accurate representation of compressible flows, even in the presence of turbulence and shock waves. 
  
\end{abstract}

\maketitle

\section{Introduction}
The development of accurate and efficient numerical methods for the simulation of compressible fluid flows remains a highly active research field in computational fluid dynamics (CFD), and is of great importance to many natural phenomena and engineering applications. Compressiblilty is usually measured by the Mach number, $Ma=u / c_s$, defined as the ratio of the flow velocity to the speed of sound and is mainly characterized by the importance of density and temperature variations and a dilatational velocity component. The presence of shock waves in compressible flows also imposes severe challenges for an accurate numerical simulation. Shock waves are sharp discontinuities of the flow properties across a thin region with the thickness of the order of mean free path.
Since in practical simulations, it is impossible to use a grid size fine enough to resolve the physical shock structure defined by the molecular viscosity, most numerical schemes rely on some numerical dissipation to stabilize the simulation and capture the shock over a few grid points \cite{caughey2003computational,vonneumann1950method}. The additional numerical dissipation of shock capturing schemes, however, is problematic in smooth turbulent regions of the flow, where a non-dissipative scheme is required to capture the complex physics accurately. Therefore, in recent years, much effort has been devoted to developing numerical schemes capable of treating shocks and turbulence, simultaneously. This has resulted in various improvements of the WENO scheme \cite{shu1999high,subramaniam2019high,fu2018new,fu2019very}, artificial diffusivity approaches \cite{haga2019robust} and hybrid schemes \cite{visbal2005shock}, to name a few.

In the past decades, the lattice Boltzmann method (LBM) has received considerable attention for the CFD as a kinetic theory approach based on the discrete Boltzmann equation. LBM has been proved to be a viable and efficient tool for the simulation of complex fluid flows and has been applied to a wide range of fluid dynamics problems including, but not limited to, turbulence \cite{dorschner2016entropic}, multi-phase flows \cite{wagner2006thermodynamic} and relativistic hydrodynamics \cite{mendoza2010fast}. The attractiveness of the LBM over conventional CFD methods, lies in the simplicity and locality of its underlying numerical algorithm which can be summarized as "stream populations along the discrete velocities $\bm{c}_i$
and equilibrate at the nodes $\bm{x}$". It is, however, well known that LBM faces stiff challenges in dealing with high-speed flows and its success has been mainly limited to low-speed incompressible flow applications.

While LBM on standard lattices 
recovers the Navier--Stokes (NS) equations in the hydrodynamic limit, there exist Galilean non-invariant error terms in the stress tensor which are negligible only in the limit of vanishing velocities and at a singular temperature, known as the lattice temperature. This prevents LBM from going to higher velocities as well as incorporating temperature dynamics. 
A natural approach to overcome this limitation is to include more discrete velocities and use the hierarchy of admissible high-order (or multi-speed) lattices \cite{shan2006kinetic,chikatamarla2009lattices} to ensure the Galilean invariance and temperature independence of the stress tensor. Although models based on high-order lattices \cite{frapolli2016entropic,wilde2020semi} have been shown to be successful in simulating compressible flows to some extent, they increase significantly the computational cost and suffer from a limited temperature range \cite{frapolli2017entropic}, as well. %

Another approach, which has received considerable attention in recent years, maintains the simplicity and efficiency 
of the standard lattices and employs correction terms in order to remove the aforementioned spurious terms in the stress tensor \cite{prasianakis2008lattice,prasianakis2009lattice}. Due to intrinsic non-uniqueness of the correction term, different implementations exist in the literature, all recover the same equations in the hydrodynamic limit \cite{guo2007thermal,feng2015three,saadat2019lattice}. See \citet{hosseini2020compressibility} for a detailed review of different implementations. Besides correction term, to fully recover the Navier-Stokes-Fourier (NSF) equations, one also needs to incorporate the energy equation. For doing that different models have been proposed in the literature which, in general, can be categorized into two main groups: hybrid and two-population methods. Hybrid methods \cite{feng2016compressible,feng2019hybrid,guo2020improved} rely on solving the total energy equation using conventional numerical schemes like finite-difference or finite-volume. However, the majority of hybrid LB schemes suffer from lack of energy conservation, as the energy equation is solved in a non-conservative form \cite{zhao2020toward}. In the two-population approach \cite{guo2007thermal,li2012coupling,karlin2013consistent,saadat2019lattice}, however, another population is used for the conservation of total energy. The latter provides a fully conservative and unified kinetic framework for the compressible flows. Previous attempts of the simulation of supersonic flows within the two-population framework on standard lattices \cite{dorschner2018particles,saadat2019lattice,saadat2020semi} have been based on the concept of shifted lattices \cite{frapolli2016lattice} or adaptive lattices \cite{dorschner2018particles} and need some form of interpolation during the streaming step. 
Therefore, a fully on-lattice conservative scheme capable of capturing the complex physics of compressible flows involving shock waves is still needed.

In this paper, we revisit and propose a two-population realization of the compressible LB model on standard lattices and investigate its accuracy and performance for a range of compressible cases from subsonic to moderately supersonic regime with shock waves and turbulence. The model is fully on-lattice and uses the single relaxation time (SRT) Bhatnagar–Gross–Krook \cite{bhatnagar1954model} (BGK) collision term along with the product-form formulation \cite{karlin2010factorization} of the equilibrium populations with a consistent correction term that restores the correct stress tensor. Through Chapman-Enskog analysis, the model recovers the compressible NSF equations with adjustable Prandtl number and adiabatic exponent in the hydrodynamic limit. It is shown that the model can accurately simulate compressible flows. Moreover, computing the correction terms with a simple upwind scheme provides enough numerical dissipation to avoid the Gibbs oscillations, and effectively capture the shock waves without degrading the accuracy of the scheme and overwhelming the physical dissipation in smooth regions. This is demonstrated through simulation of acoustic waves in the shock-vortex interaction problem. We then investigate a more challenging case of compressible decaying isotropic turbulence at large turbulent Mach numbers and Reynolds number, where interaction of compressibility effects, turbulence and shocks are present in the flow field.

The remainder of the paper is organized as follows: The kinetic equations of the two-population compressible LB model along with the pertinent equilibrium and quasi-equilibrium populations are presented in Sec.\ \ref{Sec:Kinetic_eqns}. In  Sec.\ \ref{Sec:Results}, the model is validated and analyzed through simulation of benchmark test-cases, including Sod shock-tube, shock-vortex interaction and decaying of a compressible isotropic turbulence. Conclusions are drawn in Sec.\ \ref{Sec:Conclusion}.

\section{Model description} \label{Sec:Kinetic_eqns}
\subsection{Kinetic equations}
In the two-population approach, conservation laws are split between the two sets. A set of $f$-populations $f_i$ represents mass and  momentum while another set of $g$-populations $g_i$ is earmarked for the energy conservation. Following \citet{karlin2013consistent}, we consider a single relaxation time lattice Bhatnagar--Gross--Krook (LBGK) equations for the $f$-populations and a quasi-equilibrium LBM equation for the $g$-populations, corresponding to discrete velocities $\bm{c}_i$, where $i=0,\dots,Q-1$,
\begin{align}
    f_i(\bm{x}+\bm{c}_i\delta t,t+\delta t) -f_i(\bm{x},t) &=  \omega (f_i^{\rm ex} -f_i), \label{eq:fLBGK} \\ 
    g_i(\bm{x}+\bm{c}_i\delta t,t+\delta t) -g_i(\bm{x},t) &=  \omega_1 (g_i^{\rm eq} -g_i) \nonumber\\
    	&+ (\omega - \omega_1)(g_i^{\rm eq} -g_i^{*}). \label{eq:gLBGK}
\end{align}
The extended equilibrium $f_i^{\rm ex}$, the equilibrium $g_i^{\rm eq}$ and the quasi-equilibrium $g_i^*$ satisfy the local conservation laws for the density $\rho$, momentum $\rho \bm{u}$ and energy $\rho E$,
\begin{align}
    \rho &= \sum_{i=0}^{Q-1} f_i^{\rm {ex}} = \sum_{i=0}^{Q-1} f_i, \label{eq:density} \\
	\rho \bm{u} &= \sum_{i=0}^{Q-1} \bm{c}_i f_i^{\rm {ex}} = \sum_{i=0}^{Q-1} \bm{c}_i f_i, \label{eq:momentum} \\
	\rho E &= \sum_{i=0}^{Q-1} g_i^{\rm eq} = \sum_{i=0}^{Q-1} g_i^{*} = \sum_{i=0}^{Q-1} g_i. \label{eq:energy}
\end{align}
We consider a general caloric equation of state of ideal gas. Without loss of generality, the reference temperature is set at $T=0$ and the internal energy at unit density $U$ is written as,
	\begin{align}
		U=\int_{0}^T C_v(T) dT, \label{eq:U}
	\end{align}
where $T$ is the temperature and $C_v(T)$ is the mass-based specific heat at constant volume.
The energy at unit density $E$ is,
\begin{align}
	E=U + \frac{u^2}{2}. \label{eq:E}
\end{align}
The relaxation parameters $\omega$ and $\omega_1$ are related to viscosity 
and thermal conductivity, as it will be shown below. 
We now proceed with specifying the equilibria and quasi-equilibria for the standard lattice.

\subsection{Discrete velocities and factorization}
We consider the $D3Q27$ set of three-dimensional discrete velocities $\bm{c}_i$, where $D=3$ is the space dimension and $Q=27$ is the number of discrete speeds,
\begin{equation}
	\bm{c}_i=(c_{ix},c_{iy},c_{iz}),\ c_{i\alpha}\in\{-1,0,1\},\ i=0,\dots, 26. 
	\label{eq:d3q27vel}
\end{equation}
Below, we make use of a product-form to represent all pertinent populations, the extended $f$-equilibrium, and the $g$-equilibrium and $g$-quasi-equilibrium, featured in the relaxation terms of (\ref{eq:fLBGK}) and (\ref{eq:gLBGK}). 
We follow \citet{karlin2010factorization} and 
consider a triplet of functions in two variables $\xi$ and $\PP$,
\begin{align}
	\Psi_{0}(\xi,\PP) &= 1 - \PP, 
	\label{eqn:phi0S}
	\\
	\Psi_{1}(\xi, \PP) &= \frac{1}{2}\left(\xi+ \PP\right),
	\label{eqn:phiPlusS}
	\\
	\Psi_{-1}(\xi,\PP) &= \frac{1}{2}\left(-\xi + \PP\right).
	\label{eqn:phiMinusS}
\end{align}
For vector-parameters $(\xi_x,\xi_y,\xi_z)$ and $(\PP_{xx},\PP_{yy},\PP_{zz})$,
we consider a product associated with the speeds $\bm{c}_i$ (\ref{eq:d3q27vel}),
\begin{align}
   \Psi_i= \Psi_{c_{ix}}(\xi_x,\PP_{xx}) \Psi_{c_{iy}}(\xi_y,\PP_{yy}) \Psi_{c_{iz}}(\xi_z,\PP_{zz}).
    \label{eq:productform}
\end{align}
The  moments of the product-form (\ref{eq:productform}),
\begin{equation}
	\M_{lmn}=\sum_{i=0}^{26}c_{ix}^l c_{iy}^m c_{iz}^n \Psi_i,
\end{equation}
are readily computed thanks to the factorization,
\begin{equation}
	\M_{lmn}=\M_{l00}\M_{0m0}\M_{00n}, 		\label{eq:mom27UniQuE}
\end{equation}
where $\M_{000}=1$, and where
\begin{align}
	\M_{l00}&=\left\{\begin{aligned}
		&	\xi_{x}, & l\ \text{odd}\\
		&	\PP_{xx}, & l\ \text{even}
	\end{aligned}
	\right.,
	\\
	\M_{0m0}&=\left\{\begin{aligned}
		&	\xi_{y}, & m\ \text{odd}\\
		&   \PP_{yy}, & m\ \text{even}
	\end{aligned}
	\right.,
	\\
	\M_{00n}&=\left\{\begin{aligned}
		&	\xi_{z}, & n\ \text{odd}\\
		&	\PP_{zz}, & n\ \text{even}
	\end{aligned}
	\right..
\end{align}
With the product-form (\ref{eq:productform}), we proceed to specifying the extended equilibrium $f$-populations $f^{\rm ex}_i$ in (\ref{eq:fLBGK}), and the equilibrium $g$-populations $g^{\rm eq}_i$ and the quasi-equilibrium $g$-populations $g_i^{*}$ in (\ref{eq:gLBGK}).

\subsection{Extended $f$-equilibrium}

The extended equilibrium featured in the LBGK equation (\ref{eq:fLBGK}) has been already introduced by \citet{saadat2021extended} for the fixed temperature case.
We shall summarize the construction for the purpose of the present compressible flow situation. 
At first, we define the equilibrium $f_i^{\rm eq}$ by specifying,
\begin{align}
\xi_\alpha &= u_\alpha, \label{eq:ci_f} \\ 
\PP_{\alpha\alpha}^{\rm eq} &= RT + u_\alpha^2.  \label{eq:Pi_f}
\end{align}
Substituting (\ref{eq:ci_f}) and (\ref{eq:Pi_f}) into (\ref{eq:productform}), we obtain,
\begin{align}
f_i^{\rm eq} = \rho \Psi_{c_{ix}}(u_x,\mathcal{P}_{xx}^{\rm eq}) \Psi_{c_{iy}}(u_y,\mathcal{P}_{yy}^{\rm eq}) \Psi_{c_{iz}}(u_z,\mathcal{P}_{zz}^{\rm eq}).
\label{eq:27eq}
\end{align}
The factorization (\ref{eq:mom27UniQuE}) implies that equilibrium (\ref{eq:27eq}) verifies the maximal number $Q=27$ of the moment relations established by the Maxwell--Boltzmann (MB) distribution, 
\begin{align}
	\sum_{i=0}^{26}c_{ix}^l c_{iy}^m c_{iz}^n f_i^{\rm eq} = 	 F_{lmn}^{\rm MB},\ 	 \text{    }   l,m,n \in \{0,1,2\}, \label{eq:MBmoments27}
\end{align}
where
\begin{align}
F_{lmn}^{\rm MB}=	\rho (2 \pi R T)^{-\frac{3}{2}} \int c_x^l c_y^m c_z^n e^{-\frac{(\bm{c} - \bm{u})^2}{2RT}} d\bm{c}.
\end{align}
Furthermore, with (\ref{eq:mom27UniQuE}), we find the pressure tensor and the third-order moment tensor at the equilibrium (\ref{eq:27eq}),
\begin{align}
	\bm{P}^{\rm eq} &= \sum\limits_{i = 0}^{26} \bm{c}_{i}\otimes\bm{c}_{i}f_i^{\rm eq} = \bm{P}^{\rm MB}, \label{eqn:Peq}\\
	\bm{Q}^{\rm eq} &=\sum\limits_{i = 0}^{26} \bm{c}_{i}\otimes\bm{c}_{i}\otimes\bm{c}_{i}f_i^{\rm eq}=\bm{Q}^{\rm MB}+\tilde{\bm{Q}}. \label{eqn:Qeq}
\end{align}
Here, the isotropic parts, $\bm{P}^{\rm MB}$ and $\bm{Q}^{\rm MB}$, are the Maxwell--Boltzmann pressure tensor and the third-order moment tensor, respectively,
\begin{align}
	\bm{P}^{\rm MB}&=P\bm{I}+\rho \bm{u}\otimes\bm{u},\label{eqn:PMB}\\
	\bm{Q}^{\rm MB}&=  {\rm sym}(P\bm{I}\otimes\bm{u})+\rho \bm{u}\otimes\bm{u}\otimes\bm{u},
	\label{eqn:QMB} 
\end{align}
where $P=\rho RT$ is the pressure, ${\rm sym}(\dots)$ denotes symmetrization and $\bm{I}$ is the unit tensor.

The anisotropy of the equilibrium (\ref{eq:27eq}) manifests with the deviation $\tilde{\bm{Q}} =\bm{Q}^{\rm eq}- \bm{Q}^{\rm MB}$ in (\ref{eqn:Qeq}), where only the diagonal elements are non-vanishing,
\begin{align}
	\tilde{{Q}}_{\alpha\beta\gamma}= 
	\left\{\begin{aligned}
		&\rho u_{\alpha}(1-3 RT)-\rho u_{\alpha}^3, &\text{ if }\alpha = \beta = \gamma, & \\ 
		&0,                                         & \text{otherwise}.                  &\\
	\end{aligned}\right. 
	\label{eqn:Deviation}
\end{align}
The origin of the diagonal anomaly (\ref{eqn:Deviation}) is the geometric constraint featured by the discrete speeds (\ref{eq:d3q27vel}), $c_{i\alpha}^3 =  c_{i\alpha}$, for any $i=0,\dots,26$. Put differently, the equilibrium pressure tensor (\ref{eqn:Peq}) and the off-diagonal elements of the equilibrium third-order moments (\ref{eqn:Qeq}) are included in the set of independent moments (\ref{eq:MBmoments27}), hence they verify the Maxwell--Boltzmann moment relations by the product-form. Contrary to that, the diagonal components $Q^{\rm eq}_{\alpha\alpha\alpha}$ are not among the set of moments (\ref{eq:d3q27vel}), hence the anomaly.
A remedy, commonly employed in the conventional LBM for incompressible flow simulations, is to minimize the spurious effects of the said anisotropy by fixing the lattice reference temperature, $RT_L={1}/{3}$ in order to eliminate the linear term $O(u_{\alpha})$ in  (\ref{eqn:Deviation}). Thus, the use of the equilibrium (\ref{eq:27eq}) in the LBGK equation (\ref{eq:fLBGK}) imposes a two-fold restriction: the temperature cannot be chosen differently from $T_L$ while at the same time the flow velocity has to be maintained asymptotically vanishing. While the equilibrium (\ref{eq:27eq}) can still be used for the thermal LBM in the Bussinesq approximation \cite{karlin2013consistent}, they make (\ref{eq:27eq}) insufficient for a compressible flow setting.

Instead, as was proposed by \citet{saadat2021extended}, the equilibrium (\ref{eq:27eq}) needs to be {\it extended} in such a way 
that the third-order moment anomaly (\ref{eqn:Deviation}) is compensated in the hydrodynamic limit.
Because the anomaly only concerns the diagonal elements of the third-order moments, the cancellation is achieved by redefining the diagonal elements of the second-order moments $\PP_{\alpha\alpha}$. 
As was demonstrated in \cite{saadat2021extended}, in order the achieve cancellation of the errors,  
the diagonal elements $\PP_{\alpha\alpha}^{\rm ex}$ must be extended as
\begin{align}
\mathcal{P}_{\alpha\alpha}^{\rm ex}&=\mathcal{P}_{\alpha\alpha}^{\rm eq}
	+ \delta t\left(\frac{2-\omega}{2\rho\omega}\right)\partial_\alpha \tilde{Q}_{\alpha\alpha\alpha},
	\label{eq:Pa_star}
\end{align}
where  $\partial_{\alpha}=\partial/\partial x_{\alpha}$ and $\tilde{{Q}}_{\alpha\alpha\alpha}$ is the diagonal element of the anomaly (\ref{eqn:Deviation}),
\begin{align}
	\tilde{{Q}}_{\alpha\alpha\alpha}&=\rho u_{\alpha}(1-3 RT)-\rho u_{\alpha}^3.\label{eq:anomalya}
\end{align}
With (\ref{eq:Pa_star}) instead of (\ref{eq:Pi_f}), the extended equilibrium $f_i^{\rm ex}$ is defined using the product form as before,
\begin{align}
f_i^{\rm ex} = \rho \Psi_{c_{ix}}(u_x,\mathcal{P}_{xx}^{\rm ex}) \Psi_{c_{iy}}(u_y,\mathcal{P}_{yy}^{\rm ex}) \Psi_{c_{iz}}(u_z,\mathcal{P}_{zz}^{\rm ex}).
\label{eq:extf}
\end{align}
The pressure tensor of the extended equilibrium is thus
\begin{align}
	\bm{P}^{\rm ex}=\bm{P}^{\rm eq}+\delta t \left(\frac{2-\omega}{2\omega}\right)\bgrad\cdot\tilde{\bm{Q}}.\label{eq:Pstar}
\end{align}
As it has been shown in \cite{saadat2021extended}, when the extended equilibrium (\ref{eq:extf}) is used in the LBGK equation (\ref{eq:fLBGK}) at a {\it fixed} temperature $T$, the Navier--Stokes equation for the flow momentum is recovered in a range of flow velocities and temperatures. However, in the problem under consideration, the temperature is input from the $g$-population dynamics, specifically, by solving the integral equation (\ref{eq:E}). We thus turn our attention to specifying the equilibrium and the quasi-equilibrium in the $g$-kinetic equation (\ref{eq:gLBGK}).

\subsection{$g$-equilibrium and $g$-quasi-equilibrium}

We first consider the moments of the Maxwell--Boltzmann energy distribution function,
\begin{align}
    G_{lmn}^{\rm MB} = \rho (2 \pi R T)^{-\frac{3}{2}} \int c_x^l c_y^m c_z^n  \left( \frac{c^2}{2} e^{-\frac{(\bm{c} - \bm{u})^2}{2RT}}\right) d\bm{c}.
    \label{eq:MBE}
\end{align}
Let us introduce operators  $\mathcal{O}_\alpha$ acting on any smooth function $A(\bm{u},T)$ as follows \cite{karlin2013consistent},
\begin{align}
	\mathcal{O}_\alpha A = RT \frac{\partial A}{\partial u_\alpha} + u_\alpha A. \label{eq:Operator}
\end{align}
The Maxwell--Boltzmann energy moments (\ref{eq:MBE}) can be written as the result of repeated application of operators (\ref{eq:Operator}) on the generating function,
\begin{align}
    G_{lmn}^{\rm MB} = \rho\mathcal{O}_x^l\mathcal{O}_y^m \mathcal{O}_z^n E^{\rm MB}, \label{eq:MBEcompact}
\end{align}
where the generating function $E^{\rm MB}$ is the energy of the ideal monatomic gas at unit density (three translational degrees of freedom, $C_v=(3/2)R$),
\begin{align}
	E^{\rm MB}=\frac{3}{2}RT+\frac{u^2}{2}.	\label{eq:Etr}
\end{align}
Next, we extend the Maxwell--Boltzmann energy moments (\ref{eq:MBEcompact}) to a general caloric ideal gas equation of state (\ref{eq:E}). 
	This amounts to replacing the generating function (\ref{eq:Etr}) with the energy (\ref{eq:E}), 
	\begin{align}
		G_{lmn}^{\rm eq} = \rho\mathcal{O}_x^l\mathcal{O}_y^m \mathcal{O}_z^n E. \label{eq:Ecompact}
	\end{align}
Among the higher-order moments (\ref{eq:Ecompact}), we recognize those pertinent to the hydrodynamic limit of the energy equation to be analyzed below. 
These are the equilibrium energy flux $\bm{q}^{\rm eq}$ and the flux of the energy flux tensor  $\bm{R}^{\rm eq}$,
\begin{align}
 {q}^{\rm eq}_{\alpha} &= \rho\mathcal{O}_\alpha E=\left(H + \frac{u^2}{2}\right) \rho{u}_{\alpha}, \label{eq:qeq}\\
 {R}^{\rm eq}_{\alpha\beta} &=\rho \mathcal{O}_\alpha \mathcal{O}_\beta E= \left(H + \frac{u^2}{2}\right){P}^{\rm eq}_{\alpha\beta} + P {u}_{\alpha}{u}_{\beta}. \label{eq:Req}
\end{align}
Here $H$ is the specific enthalpy,
\begin{align}
	H=\int_{0}^T C_p(T) dT,
\end{align}
while $C_p$ is the specific heat at constant pressure, satisfying Mayer's relation, $C_p-C_v=R$.

The equilibrium populations $g_i^{\rm eq}$ are specified with the operator version of the product-form (\ref{eq:productform}). To that end, we consider parameters ${\xi_\alpha}$ and $\PP_{\alpha\alpha}$ as operator symbols,
\begin{align}
    \xi_\alpha &= \mathcal{O}_\alpha, \label{eq:Oa}\\
    \PP_{\alpha\alpha} &= \mathcal{O}_\alpha^2. \label{eq:OaOa}
\end{align}
With the operators (\ref{eq:Oa}) and (\ref{eq:OaOa}) substituted into the product form (\ref{eq:productform}), the equilibrium populations $g_i^{\rm eq}$ are written using the generating function (\ref{eq:E}),
\begin{align}
    g_i^{\rm eq} = \rho \Psi_{c_{ix}}(\mathcal{O}_x,\mathcal{O}_x^2) \Psi_{c_{iy}}(\mathcal{O}_y,\mathcal{O}_y^2) \Psi_{c_{iz}}(\mathcal{O}_z,\mathcal{O}_z^2)E. \label{eq:geq_i}
\end{align}
With (\ref{eq:mom27UniQuE}), it is straightforward to see that the equilibrium (\ref{eq:geq_i}) verifies a subset of the equilibrium energy moments (\ref{eq:Ecompact}),
\begin{align}
    \sum_{i=0}^{26}c_{ix}^l c_{iy}^m c_{iz}^n g_i^{\rm eq} = G_{lmn}^{\rm eq}, \text{    }   l,m,n \in \{0,1,2\}. \label{eq:geqmom}
\end{align}
Thus, by construction, the $g$-equilibrium (\ref{eq:geq_i}) recovers the maximal number $Q=27$ of the energy moments (\ref{eq:Ecompact}), including the energy flux (\ref{eq:qeq}) and the flux of the energy flux (\ref{eq:Req}).

Finally, similarly to \cite{karlin2013consistent}, the quasi-equilibrium populations $g_i^*$ are needed for adjusting the Prandtl number of the model. To that end, the quasi-equilibrium $g_i^*$ differs from $g_i^{\rm eq}$ by the non-equilibrium energy flux only,
	\begin{align}
		g_{i}^*= \left\{\begin{aligned}
			& g_{i}^{\rm eq}+\frac{1}{2}\bm{c}_i\cdot\left(\bm{q}^*-\bm{q}^{\rm eq}\right), &\text{ if } c_i^2=1, & \\ 
			&g_i^{\rm eq}, & \text{otherwise}.&\\
		\end{aligned}\right.
	\label{eq:gstar}	
	\end{align}
Here $\bm{q}^*$ is a specified quasi-equilibrium energy flux. Indeed, (\ref{eq:gstar}) and (\ref{eq:geqmom}) imply for $l,m,n \in \{0,1,2\}$,
\begin{align}
	\sum_{i=0}^{26}c_{ix}^l c_{iy}^m c_{iz}^n g_i^{*} = 
	\left\{\begin{aligned}
		& q^{*}_x, &\text{ if } l=1,m=0,n=0 & \\ 
		& q^{*}_y, &\text{ if } l=0,m=1,n=0 & \\ 
		& q^{*}_z, &\text{ if } l=0,m=0,n=1 & \\ 
		&G_{lmn}^{\rm eq}, & \text{otherwise}.&\\
	\end{aligned}\right. 
\label{eq:gqeqmom}
\end{align}
While the above construction holds for any specified $\bm{q}^*$, the quasi-equilibrium flux required for the consistent realization of the adjustable Prandtl number by the LBM system (\ref{eq:fLBGK}) and (\ref{eq:gLBGK}) reads,
\begin{align}
\bm{q}^{*} &= \bm{q}^{\rm eq} +  \bm{u}\cdot\left(\bm{P}-\bm{P}^{\rm eq}+\frac{\delta t}{2}\bgrad\cdot\tilde{\bm{Q}}\right), \label{eq:qstar1}
\end{align}
where $\bm{P}$ is the pressure tensor,
\begin{align}
	\bm{P} &= \sum\limits_{i = 0}^{26} \bm{c}_{i}\otimes\bm{c}_{i}f_i. \label{eqn:P}
\end{align}
Note that unlike in the original incompressible thermal model \cite{karlin2013consistent}, the quasi-equilibrium flux (\ref{eq:qstar1}) now includes an extension due to the diagonal anomaly. With all the elements of the LBM system (\ref{eq:fLBGK}) and (\ref{eq:gLBGK}) specified, we now proceed with working out its hydrodynamic limit.

\subsection{Hydrodynamic limit}

Taylor expansion of the shift operator in (\ref{eq:fLBGK}) and (\ref{eq:gLBGK}) to second order gives,
\begin{align}
	\left[\delta t D_i+\frac{\delta t^ 2}{2}D_iD_i\right]f_i=&\omega(f_i^{\rm ex}-f_i),\label{eq:fTaylor}\\
\left[\delta t D_i+\frac{\delta t^ 2}{2}D_iD_i\right]g_i=&\omega_1(g_i^{\rm eq}-g_i)\nonumber\\
&+(\omega - \omega_1)(g_i^{\rm eq} -g_i^{*}),\label{eq:gTaylor}
\end{align}
where $D_i$ is the derivative along the characteristics,
\begin{equation}
	D_i=\partial_t + \bm{c}_i\cdot\bgrad.
\end{equation}
Introducing a multi-scale expansion,
\begin{align}
	f_i&= f_i^{(0)} + \delta t f_i^{(1)} + \delta t^2 f_i^{(2)} + O(\delta t^3),  \\
	f_i^{\rm ex} &= f_i^{{\rm ex}(0)} + \delta t f_i^{{\rm ex}(1)} + \delta t^2 f_i^{{\rm ex}(2)} + O(\delta t^3), \\
	g_i&= f_i^{(0)} + \delta t f_i^{(1)} + \delta t^2 f_i^{(2)} + O(\delta t^3),  \\
	g_i^{*} &= g_i^{*(0)} + \delta t g_i^{*(1)} + \delta t^2 g_i^{*(2)} + O(\delta t^3), \\
	\partial_t &= \partial_t^{(1)} + \delta t\partial _t^{(2)} + O(\delta t^2), 
\end{align}
substituting into (\ref{eq:fTaylor}) and (\ref{eq:gTaylor}), and using the notation,
\begin{align}
	&D_i^{(1)}=\partial_t^{(1)}+\bm{c}_i\cdot \bgrad,
\end{align}
we obtain, from zeroth through second order in the time step $\delta t$, for the $f$-populations,
\begin{align}
	&f_i^{(0)}=f_i^{{\rm ex}(0)}=f_i^{\rm eq},\label{eq:zeroCE}\\
	&D_i^{(1)}f_i^{(0)}=-\omega \left(f_i^{(1)} - f_i^{{\rm ex}(1)}\right),\label{eq:oneCE}\\
	&\partial_t^{(2)}f_i^{(0)}+\bm{c}_i\cdot \bgrad f_i^{(1)} -\frac{\omega}{2}D_i^{(1)} \left(f_i^{(1)} - f_i^{{\rm ex}(1)}\right)\nonumber\\
	&=-\omega f_i^{(2)}+\omega f_i^{{\rm ex}(2)},\label{eq:twoCE}
\end{align}
and similarly for the $g$-populations,
\begin{align}
		&g_i^{(0)}=g_i^{*(0)}=g_i^{\rm eq},\label{eq:gzeroCE}\\
		&D_i^{(1)}g_i^{(0)}=-\omega_1 g_i^{(1)}-(\omega - \omega_1)g_i^{*(1)},\label{eq:goneCE}\\
		&\partial_t^{(2)}g_i^{(0)}+\bm{c}_i\cdot \bgrad g_i^{(1)} -\frac{\omega_1}{2}D_i^{(1)} g_i^{(1)}-\frac{\omega-\omega_1}{2}D_i^{(1)} g_i^{*(1)}\nonumber\\
		&=-\omega_1 g_i^{(2)}-(\omega -\omega_1) g_i^{*(2)}.\label{eq:gtwoCE}
	\end{align}
With (\ref{eq:zeroCE}) and (\ref{eq:gzeroCE}), the mass, momentum and energy conservation (\ref{eq:density}), (\ref{eq:momentum}) and (\ref{eq:energy}) imply the solvability conditions,
\begin{align}
	&\sum_{i=0}^{26} f_i^{{\rm ex}(k)} = \sum_{i=0}^{26} f_i^{(k)}=0,\ k=1,2\dots; \label{eq:density_orders} \\
	& \sum_{i=0}^{26} \bm{c}_i f_i^{{\rm ex}(k)} = \sum_{i=0}^{26} \bm{c}_i f_i^{(k)}=0,\ k=1,2,\dots;\label{eq:momentum_orders}\\
	& \sum_{i=0}^{26}  g_i^{*(k)} = \sum_{i=0}^{26}  g_i^{(k)}=0,\ k=1,2,\dots.\label{eq:energy_orders}
\end{align}
With the $f$-equilibrium (\ref{eq:27eq}) and the $g$-equilibrium (\ref{eq:geq_i}), while taking into account the solvability conditions (\ref{eq:density_orders}), (\ref{eq:momentum_orders}) and (\ref{eq:energy_orders}), and also making use of the equilibrium pressure tensor (\ref{eqn:Peq}) and (\ref{eqn:PMB}), and the equilibrium energy flux (\ref{eq:qeq}), the first-order kinetic equations (\ref{eq:oneCE}) and (\ref{eq:goneCE}) imply the following first-order balance equations for the density, momentum and energy,
\begin{align}
	&\partial_t^{(1)}\rho=-\bgrad\cdot (\rho\bm{u}),\label{eq:density1}\\
	&\partial_t^{(1)}(\rho\bm{u})=	-\bgrad\cdot(P\bm{I}+\rho \bm{u}\otimes\bm{u}).\label{eq:momentum1}\\
	&\partial_t^{(1)}(\rho E)=-\bgrad\cdot\bm{q}^{\rm eq}.\label{eq:energy1}
\end{align}
The first-order energy equation (\ref{eq:energy1}) can be recast into the temperature equation by virtue of (\ref{eq:density1}) and (\ref{eq:momentum1}),
	\begin{align}
		\rho C_v\partial_t^{(1)}T=-\rho C_v\bm{u}\cdot\bgrad T -P(\bgrad\cdot \bm{u}). \label{eq:dT1}
	\end{align}
Thus, to first order, the LBM recovers the compressible Euler equations for a generic ideal gas.

Moreover, the first-order constitutive relation for the nonequilibrium pressure tensor $\bm{P}^{(1)}$ is found from (\ref{eq:oneCE})
as follows, using (\ref{eqn:PMB}), (\ref{eqn:Qeq}), (\ref{eqn:QMB}) and  (\ref{eqn:Deviation}),
\begin{align}
	-\omega \bm{P}^{(1)}+\omega\bm{P}^{{\rm ex}(1)}=\partial_t^{(1)}\bm{P}^{\rm MB}+\bgrad\cdot\bm{Q}^{\rm MB}+\bgrad\cdot\tilde{\bm{Q}},
	\label{eq:const1}
\end{align}
where
\begin{align}
	&\bm{P}^{(1)}=\sum_{i=0}^{Q-1} \bm{c}_i\otimes \bm{c}_i f_i^{(1)},\\
	&\bm{P}^{{\rm ex}(1)}=\sum_{i=0}^{Q-1} \bm{c}_i\otimes \bm{c}_i f_i^{{\rm ex}(1)}.
\end{align}
Using (\ref{eq:density1}), (\ref{eq:momentum1}) and (\ref{eq:dT1}), we find in (\ref{eq:const1}),
\begin{align}
	\partial_t^{(1)}\bm{P}^{\rm MB}+\bgrad\cdot \bm{Q}^{\rm MB}=\bm{Z},
\label{eq:rate}
\end{align}
where we have introduced a short-hand notation for the total stress, including both the shear and the bulk contributions,
	\begin{align}
		\bm{Z}=&P\left(\bgrad \bm{u}+\bgrad\bm{u}^\dagger-\frac{2}{3}(\bgrad\cdot\bm{u})\bm{I}\right)\nonumber\\
		&+P\left(\frac{2}{3}-\frac{R}{C_v}\right)(\bgrad\cdot\bm{u})\bm{I},\label{eq:Sigma}
	\end{align}
and where $(\cdot)^\dagger$ denotes transposition. 
With (\ref{eq:rate}) and (\ref{eq:Sigma}), the nonequilibrium pressure tensor (\ref{eq:const1}) becomes,
\begin{align}
	\bm{P}^{(1)}=&-\frac{1}{\omega}\bm{Z}-\frac{1}{\omega}\bgrad\cdot\tilde{\bm{Q}}+ \bm{P}^{{\rm ex}(1)}.
	\label{eq:const12}
\end{align}
A comment is in order. In (\ref{eq:const12}), the first term is the conventional contribution from both the shear and the bulk stress. The second term is anomalous due to the diagonal anisotropy (\ref{eqn:Deviation}) while the third is the counter-term required to annihilate the spurious contribution in the next, second-order approximation. According to (\ref{eq:Pstar}),
\begin{align}
	\bm{P}^{{\rm ex}(1)}=\left(\frac{2-\omega}{2\omega}\right)\bgrad\cdot\tilde{\bm{Q}}.\label{eq:cancelation}
\end{align}

Similarly, the first-order constitutive relation for the nonequilibrium energy flux $\bm{q}^{(1)}$ is found from (\ref{eq:goneCE}),
\begin{align}
	-\omega_1\bm{q}^{(1)}-(\omega - \omega_1)\bm{q}^{*(1)}=\partial_t^{(1)} \bm{q}^{\rm eq} + \bgrad\cdot \bm{R}^{\rm eq}.
	\label{eq:constq1}
\end{align}
Evaluating the right hand side of (\ref{eq:constq1}) with the help of the first-order relations 
	(\ref{eq:density1}), (\ref{eq:momentum1}) and (\ref{eq:dT1}), we obtain,
\begin{align}
\partial_t^{(1)} \bm{q}^{\rm eq} + \bgrad\cdot \bm{R}^{\rm eq} = PC_p\bgrad T   + \left(\bm{u}\cdot \bm{Z}\right).
		\label{eqn:dt1q3}
\end{align}
With (\ref{eqn:dt1q3}), the nonequilibrium energy flux (\ref{eq:constq1}) becomes,
	\begin{align}
\bm{q}^{(1)}=-\frac{1}{\omega_1}PC_p\bgrad T -\frac{1}{\omega_1} \left(\bm{u}\cdot \bm{Z}\right) -\frac{\omega - \omega_1}{\omega_1}\bm{q}^{*(1)}.	\label{eq:constq12}
	\end{align}
The quasi-equilibrium energy flux $\bm{q}^{*(1)}$ is evaluated according to (\ref{eq:qstar1}) and by taking into account the first-order constitutive relation for the pressure tensor (\ref{eq:const12}),
\begin{align}
	\bm{q}^{*(1)}=\bm{u}\cdot\left(\bm{P}^{(1)}+\frac{1}{2}\bgrad\cdot\tilde{\bm{Q}}\right)
		=-\frac{1}{\omega}\left(\bm{u}\cdot\bm{Z}\right). \label{eq:qstar11}
\end{align}
We comment that the first term in the nonequilibrium energy flux (\ref{eq:constq12}) is a precursor of the Fourier law of thermal conductivity while the second and the third terms combine to the viscous heating contribution, as we shall see it below. The quasi-equilibrium flux (\ref{eq:qstar11}) is required for consistency of the viscous heating with the prescribed Prandtl number \cite{karlin2013consistent}.

With the first-order constitutive relations for the nonequilibrium fluxes (\ref{eq:const12}) and (\ref{eq:constq12}) in place, we proceed to the second-order approximation.
Applying the solvability condition (\ref{eq:density_orders}) and (\ref{eq:momentum_orders}) to the second-order $f$-equation (\ref{eq:twoCE}), we obtain,
\begin{align}
	&\partial_t^{(2)}\rho=0,\label{eq:rho2}\\
	&\partial_t^{(2)}(\rho\bm{u})=-\bgrad\cdot \left[\left(1-\frac{\omega}{2}\right)\bm{P}^{(1)}+\frac{\omega}{2}\bm{P}^{{\rm ex}(1)}
	\right].\label{eq:u2}
\end{align}
The second-order momentum equation (\ref{eq:u2}) is transformed by virtue of (\ref{eq:const12}) and (\ref{eq:cancelation}) to give,
\begin{align}
	\partial_t^{(2)} (\rho \bm{u})=  -\bgrad\cdot\left[-\left(\frac{1}{\omega}-\frac{1}{2}\right)\bm{Z}\right].
	\label{eq:momentum2}
\end{align}
Note that, the anomalous terms cancel out and the result (\ref{eq:momentum2}) is manifestly isotropic.

Finally, applying solvability condition (\ref{eq:energy_orders}) to the second-order $g$-equation (\ref{eq:gtwoCE}), we find
\begin{align}
	\partial_t^{(2)}(\rho E)=-\bgrad\cdot \left[\left(1-\frac{\omega_1}{2}\right)\bm{q}^{(1)}-\frac{\omega-\omega_1}{2}\bm{q}^{*(1)}\right].
	\label{eq:energy2}
\end{align}
Taking into account  the first-order energy flux (\ref{eq:constq12}) and the quasi-equilibrium energy flux (\ref{eq:qstar11}), we obtain in (\ref{eq:energy2}),
\begin{align}
	\partial_t^{(2)}(\rho E)=-\bgrad\cdot \left[-\left(\frac{1}{\omega_1}-\frac{1}{2}\right)C_p P\bgrad T\right]\nonumber\\
	-\bgrad\cdot \left[-\left(\frac{1}{\omega}-\frac{1}{2}\right)\left(\bm{u}\cdot\bm{Z}\right)\right].
	\label{eq:energy23}
\end{align}
While the first term leads to the Fourier law, it is important to note that the second term represents viscous heating consistent with the momentum equation (\ref{eq:momentum2}). The latter consistency is implied by the construction of the quasi-equilibrium energy flux (\ref{eq:qstar1}) and (\ref{eq:qstar11}).   
This concludes the second-order accurate analysis of the hydrodynamic limit of the LBM system (\ref{eq:fLBGK}) and (\ref{eq:gLBGK}), and we proceed with a summary of the gas dynamics equations thereby recovered.

\subsection{Equations of gas dynamics}

Combining the first- and second-order contributions to the density, the momentum and the energy equation, (\ref{eq:density1}) and (\ref{eq:rho2}), (\ref{eq:momentum1}) and (\ref{eq:momentum2}), and (\ref{eq:energy1}) and (\ref{eq:energy23}), respectively, and using a notation, $\partial_t=\partial_t^{(1)}+\delta t\partial_t^{(2)}$, we arrive at the continuity, the flow and the energy equations of gas dynamics as follows,
\begin{align}
    &\partial_t\rho + \bgrad\cdot(\rho\bm{u}) = 0,\label{eq:continuity} 
    \\
    &\partial_t (\rho \bm{u})
    + \bgrad \cdot (\rho \bm{u}\otimes\bm{u}) + \bgrad\cdot \bm{\pi} = 0,
	\label{eq:flow}
	\\
	&\partial_t (\rho E) + \bgrad\cdot(\rho E \bm{u})
    + \bgrad \cdot \bm{q} + \bgrad\cdot (\bm{\pi}\cdot\bm{u}) = 0. \label{eq:energyeq}
\end{align}
Here, $\bm{\pi}$ is the pressure tensor
\begin{align}
    \bm{\pi} = P\bm{I} - \mu \left(\bm{S} - \frac{2}{3}(\bgrad\cdot\bm{u})\bm{I} \right) - \varsigma(\bgrad\cdot\bm{u})\bm{I}, 
\end{align}
with $P$ the pressure of ideal gas,
\begin{align}
	P=\rho RT,
\end{align}
with the strain rate tensor
\begin{align}
    \bm{S} = \bgrad \bm{u} + \bgrad \bm{u}^\dagger, \label{eq:strain}
\end{align}
and the dynamic viscosity $\mu$ and the bulk viscosity $\varsigma$,
\begin{align}
    \mu & = \left( \frac{1}{\omega} - \frac{1}{2} \right)P\delta t, \label{eq:mu}\\
    \varsigma & = \left(\frac{2}{3}-\frac{R}{C_v}\right) \mu.\label{eq:bulkvisc}
\end{align}
The heat flux $\bm{q}$ in the energy equation (\ref{eq:energyeq}) reads
\begin{align}
    \bm{q} = -\kappa \bgrad T,
\end{align}
with the thermal conductivity coefficient $\kappa$,
\begin{align}
	\kappa &= \left(\frac{1}{\omega_1}-\frac{1}{2} \right)C_p P \delta t.\label{eq:kappa}
\end{align}
The Prandtl number due to (\ref{eq:mu}) and (\ref{eq:kappa}) is,
	\begin{equation}
		{\rm Pr}=\frac{C_p\mu}{\kappa}=\frac{\omega_1(2-\omega)}{\omega(2-\omega_1)},\label{eq:Pr}
	\end{equation}
while the adiabatic exponent, 
	 \begin{equation}
	 	\gamma=\frac{C_p}{C_v},\label{eq:gamma}
	 \end{equation}
is defined by the choice of the caloric equations of state (\ref{eq:U}) and Mayer's relation, $C_p-C_v=R$.
The mass, momentum and energy equations, (\ref{eq:continuity}), (\ref{eq:flow}) and (\ref{eq:energyeq}) are the standard equations of the macroscopic gas dynamics. We shall conclude the model development with a summary of the key elements of the LBM system (\ref{eq:fLBGK}) and (\ref{eq:gLBGK}).

\subsection{Summary of the lattice Boltzmann model}
The two-population lattice Boltzmann model (\ref{eq:fLBGK}) and (\ref{eq:gLBGK}) on the standard $D3Q27$ discrete velocity set introduced by \citet{karlin2013consistent} is extended to the compressible flow simulation following the three key modifications:
\begin{itemize}
	\item The product-form extended equilibrium for the momentum lattice, Eqs.\ (\ref{eq:productform}), (\ref{eq:extf}), (\ref{eq:Pa_star}):
	\begin{align*}
		&f_i^{\rm ex} = \rho \Psi_{c_{ix}}(u_x,\mathcal{P}_{xx}^{\rm ex}) \Psi_{c_{iy}}(u_y,\mathcal{P}_{yy}^{\rm ex}) \Psi_{c_{iz}}(u_z,\mathcal{P}_{zz}^{\rm ex});
	\end{align*}
	\item The operator product-form equilibrium for the energy lattice,
	Eqs.\ (\ref{eq:E}), (\ref{eq:Operator}), (\ref{eq:productform}), (\ref{eq:geq_i}):
\begin{align*}
&g_i^{\rm eq} = \rho \Psi_{c_{ix}}(\mathcal{O}_x,\mathcal{O}_x^2) \Psi_{c_{iy}}(\mathcal{O}_y,\mathcal{O}_y^2) \Psi_{c_{iz}}(\mathcal{O}_z,\mathcal{O}_z^2)E;
\end{align*}	
	\item The quasi-equilibrium for the energy lattice is made consistent with both of the above, Eqs.\  (\ref{eq:gstar}), (\ref{eq:qstar1}):
	\begin{align*}
	&	g_{i}^*= \left\{\begin{aligned}
		& g_{i}^{\rm eq}+\frac{1}{2}\bm{c}_i\cdot\left(\bm{q}^*-\bm{q}^{\rm eq}\right)
		 &\text{ if } c_i^2=1, & \\ 
		&g_i^{\rm eq}, & \text{otherwise}.&\\
	\end{aligned}\right.\\
	\end{align*}	
\end{itemize}
We shall proceed with the implementation of the compressible lattice Boltzmann model and numerical validation.

\section{Numerical results} \label{Sec:Results}
\subsection{General implementation issues}\label{sec:Implementation}
The spatial discretization of the deviation $\bm{\tilde{Q}}$ in Eqs.(\ref{eq:Pstar}) and (\ref{eq:qstar1}) has important effect on stability of the model, especially in the case of supersonic flows where discontinuities emerge in the flow field. It has been shown through linear stability analysis \cite{hosseini2020compressibility} that, while second-order central difference scheme provides good stability domain in the subsonic regime, the first-order upwind scheme is necessary for maintaining the stability in the supersonic regime and capturing shock wave. We, therefore, employ the first-order upwind scheme in order to have a wider stability domain.

For example, the $x$-derivative of the deviation $\tilde{Q}_{xxx}$ at grid point $\bm{x}_{i,j,k}$ can be written as  

\begin{align}
    \partial_x \tilde{Q}_{xxx,(i,j,k)} = \frac{\tilde{Q}_{xxx,(i+1/2,j,k)} - \tilde{Q}_{xxx,(i-1/2,j,k)}}{\Delta x},
\end{align}
where (omitting $xxx$ and $j,k$ indices) $\tilde{Q}_{i+1/2}$ and $\tilde{Q}_{i-1/2}$ are upwind reconstruction of $\tilde{Q}$ at the interface $\bm{x}_{i \pm 1/2,j,k}$,
\begin{align}
    	\tilde{Q}_{i+1/2} &= 
	\left\{\begin{aligned}
		&\tilde{Q}_i, &\text{ if } u_x>0, & \\ 
		&\tilde{Q}_{i+1},                                         & \text{otherwise},                  &\\
	\end{aligned}\right. 
\end{align}

The performance and accuracy of the proposed LBM for  compressible flow is tested numerically through the simulation of benchmark cases. First, the Sod shock tube problem is considered. Second, we show the ability of the model in capturing moderately supersonic shock waves, through simulation of shock-vortex interaction. Finally, the model is tested with a compressible turbulence problem, i.e, decaying of compressible homogeneous isotropic turbulence at different turbulent Mach numbers.
All simulations are performed assuming constant specific heats with gas constant $R=1$, adiabatic exponent $\gamma = 1.4$ and the $D3Q27$ lattice.

\subsection{Sod's shock tube}
\begin{figure}[]
	\includegraphics[width=0.5\textwidth]{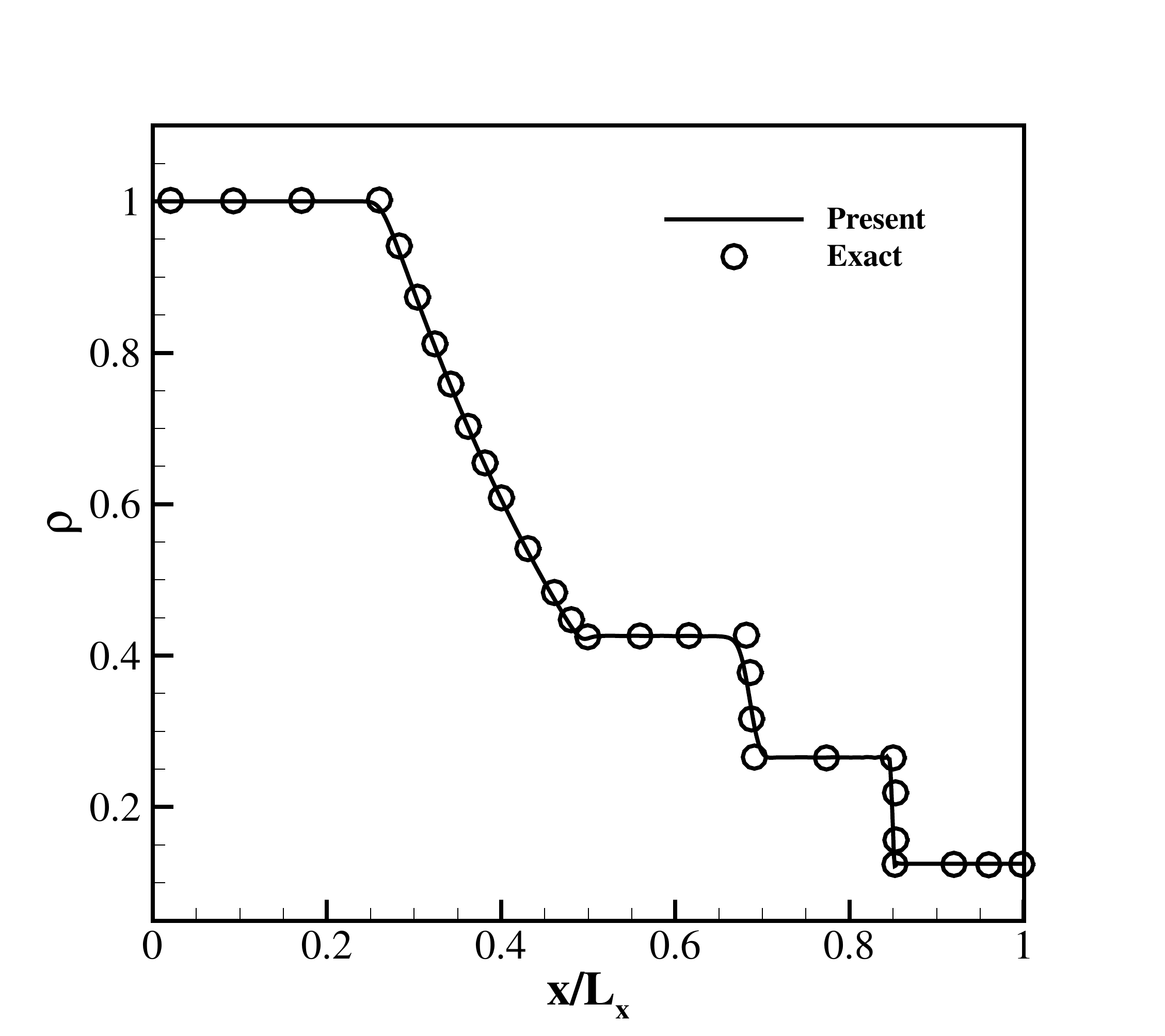}
	\caption{Density profile for Sod's shock tube simulation at non-dimensional time $t^* = 0.2$. Symbols: present model; line: exact solution.}
	\label{fig:SodRho}
\end{figure}

\begin{figure}[]
		\includegraphics[width=0.5\textwidth]{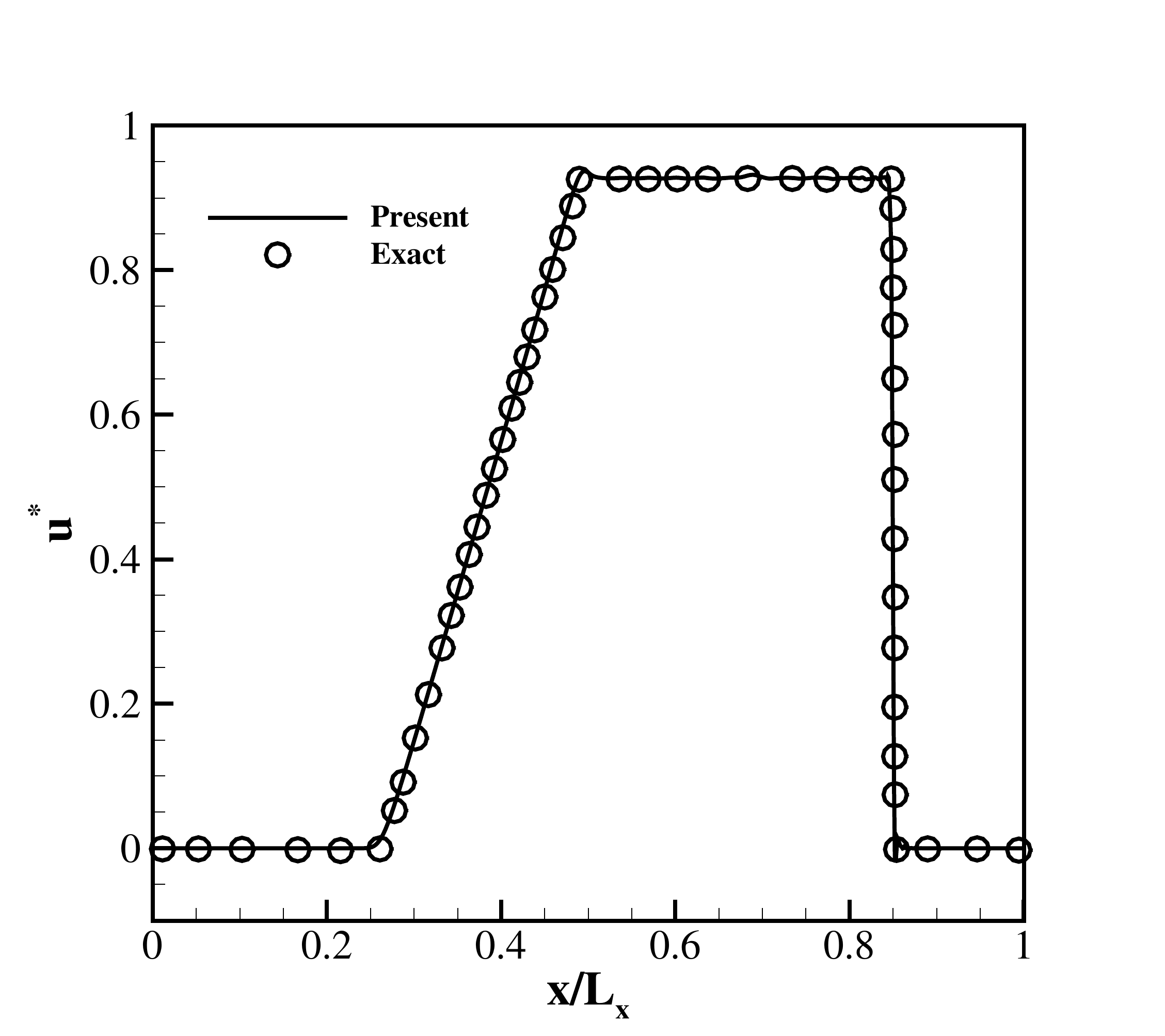}
		\caption{Reduced velocity profile for Sod's shock tube simulation at non-dimensional time $t^* = 0.2$. Symbols: present model; line: exact solution.}
		\label{fig:Sodu}
\end{figure}

Sod's shock tube benchmark \cite{sod1978survey} is a classical Riemann problem, which is often used to test capability of a compressible flow solver in capturing shock waves, contact discontinuities and expansion fans. The initial flow field is given by
\begin{equation*}
	(\rho ,u_x,P) = \left\{ \begin{aligned}
&	(1.0,0,0.15),\ &{\rm{   }}x/L_x \le 0.5,\\
&	(0.125,0,0.015),\ &{\rm{ }}x/L_x > 0.5,
	\end{aligned} \right.
\end{equation*}   
where $L_x=600$ is the number of grid points. Simulation results with the viscosity $\mu = 0.015$ for the density and reduced velocity $u^*=u/\sqrt{T_l}$, where  $T_l$ is temperature on the left half of tube, at non-dimensional time $t^*= t \sqrt{T_l} / L_x = 0.2$, are shown in Figs.\ \ref{fig:SodRho} and \ref{fig:Sodu}. It can be seen that, apart from a small oscillation, the results match the non-viscous exact solution  well.

\subsection{Shock-vortex interaction}
Sound generation by a vortex passing through a shock wave \cite{inoue1999sound} is studied to assess the performance and accuracy of the developed model for supersonic flows involving shock. This problem consists of an isentropic vortex, with vortex Mach number ${\rm Ma}_v$, initially in the upstream shock region, which is passed through a stationary shock wave at advection Mach number ${\rm Ma}_a=1.2$ with the left state $(\rho,T,u_x,u_y)_{l}=(1,0.05, {\rm Ma}_a \sqrt{\gamma T_l},0)$ and Rankine-Hugoniot right state. The initial field with standing shock $(\rho_{\infty},P_{\infty}, {u}_{x,\infty}, {u}_{y,\infty})$ is perturbed with an isentropic vortex with radius $r_v$ centered at $(x_v,y_v)$ \cite{inoue1999sound}
\begin{align*}
    \rho &= \rho_\infty {\left[ {1 - \frac{{\gamma  - 1}}{2}{\rm Ma}_v^2{e^{(1 - {r^2})}}} \right]^{1/\left( {\gamma  - 1} \right)}}, \\
    P &= P_\infty {\left[ {1 - \frac{{\gamma  - 1}}{2}{\rm Ma}_v^2{e^{(1 - {r^2})}}} \right]^{\gamma/\left( {\gamma  - 1} \right)}}, \\
    u_x &= u_{x,\infty} + \sqrt {\gamma {T_l}} {\rm Ma}_v\frac{{\left( {y - {y_v}} \right)}}{{{r_v}}}{e^{(1 - {r^2})/2}}, \\ 
	u_y &=u_{y,\infty} -\sqrt {\gamma {T_l}} {\rm Ma}_v\frac{{\left( {x - {x_v}} \right)}}{{{r_v}}}{e^{(1 - {r^2})/2}},
\end{align*}
where $r = \sqrt {{{\left( {x - {x_v}} \right)}^2} + {{\left( {y - {y_v}} \right)}^2}} / r_v$ is the reduced radius and the shock is initially located at $x_s = 8 r_v$.
\begin{figure}[t]
		\centering
		\includegraphics[width=0.5\textwidth]{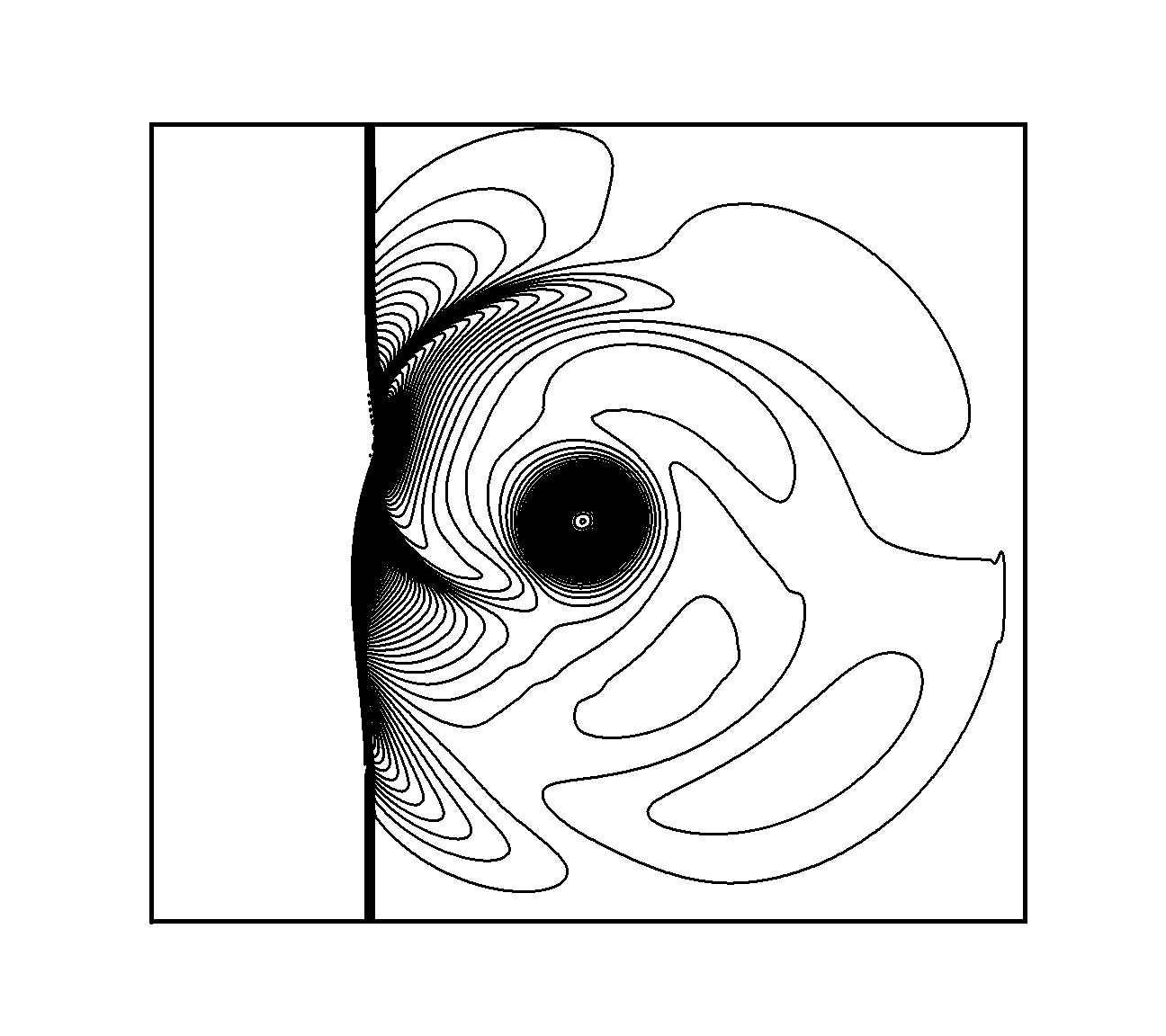}
		\caption{The sound pressure field $\Delta P$ for the shock-vortex interaction with ${\rm Ma}_a=1.2$, ${\rm Ma}_v=0.25$ and ${\rm Re}=800$ at $t^*=6$. The contour levels are from $\Delta P_{min}=-0.48$ to $\Delta P_{max}=0.16$ with an increment of $0.003216$.
}
		\label{fig:SoundContour}
\end{figure} 

We perform a simulation with ${\rm Ma}_a = 1.2$, ${\rm Ma}_v=0.25$, where the Reynolds number is set to ${\rm Re} = \frac{\rho_L c_{s,l} r_v}{\mu} = 800$, $c_{s,l}$ is the speed of sound upstream of the shock, and the Prandtl number is ${\rm Pr} = 0.75$.
The computational domain size is $L_x \times L_y=1680\times1440$, the vortex radius is $r_v = L_x/28$ and the vortex center is at $(x_v,y_v)=(6 r_v, L_y/2)$. 

\begin{figure}[]
		\centering
		\includegraphics[width=0.5\textwidth]{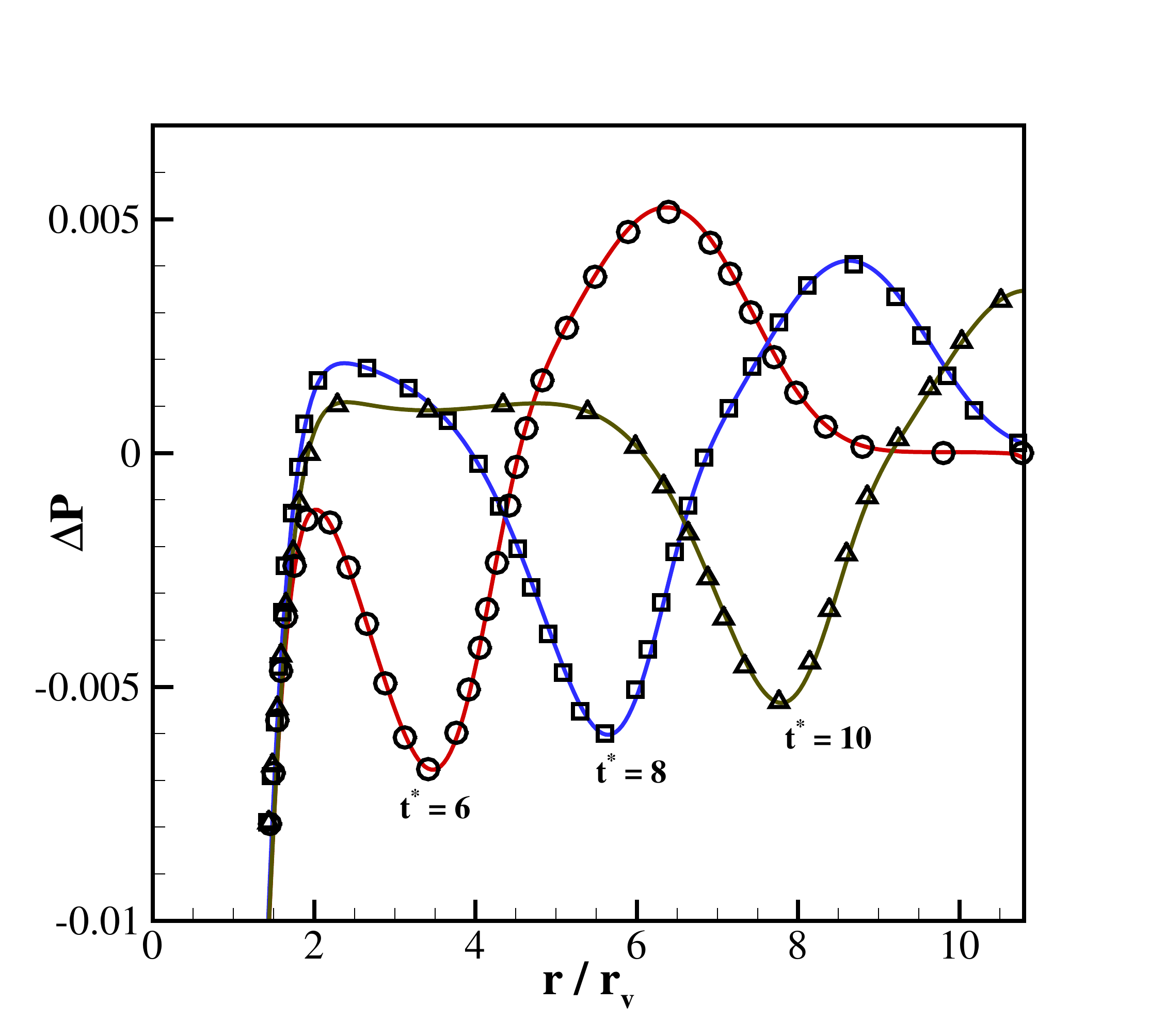}
		\caption{Comparison of radial sound pressure distribution $\Delta P$ for ${\rm Ma}_a = 1.2$, ${\rm Ma}_v = 0.25$ and ${\rm Re} = 800$ with the DNS results at three different times $t^* = 6, 8, 10$. Lines: present model; symbol: DNS \cite{inoue1999sound}.}
		\label{fig:SoundComparison}
\end{figure}

Fig.\ \ref{fig:SoundContour} shows the sound pressure contours at time $t^*=6$, where the sound pressure is defined as, $\Delta  P = (P-P_s)/P_s$, 
and $P_s$ is the pressure behind the shock wave. The shock wave deformation caused by the interaction with the vortex is observed. To quantify the accuracy of the computations, the radial sound pressure distribution is plotted in Fig. \ref{fig:SoundComparison}  in comparison with the DNS results \cite{inoue1999sound} . The sound pressure is measured in the radial direction with the origin at the vortex center, at an angle $\theta = -45^\circ$ and at three different non-dimensional times $t^*=6, 8, 10$, where $t^*=t{c_{s,l}}/{r_v}$. Excellent agreement is observed between the present model and the DNS \cite{inoue1999sound}. Note that the sound pressure is typically a small perturbation (around 1\%) on top of the hydrodynamic pressure. This shows that the present model with the LBGK collision term can accurately capture moderately supersonic shock waves. 

\subsection{Decaying of compressible isotropic turbulence}
To demonstrate that the present compressible model is a reliable method for the simulation of complex flows involving both turbulence and shocks, decaying compressible homogeneous isotropic turbulence in a periodic box is considered as the final test-case. This problem has been studied extensively \cite{lee1991eddy,mansour1994decay,samtaney2001direct,johnsen2010assessment,kumar2013weno,cao2019three, chen2020simulation} and is a challenging test-case, as it contains both compressibility effects and shocks, as well as turbulent structures in the flow field \cite{johnsen2010assessment}. 

The initial condition in a cubic domain with the side $L$ is set to be unit density and constant temperature along with a divergence-free velocity field which follows the specified energy spectrum,
\begin{align}
      \mathcal{E}(\kappa) &= A \kappa^4 exp\left(-2(\kappa/\kappa_0)^2\right),
\end{align}
where $\kappa$ is the wave number, $\kappa_0$ is the wave number at which the spectrum peaks and the amplitude $A$ controls the initial kinetic energy \cite{samtaney2001direct}. The method of kinematic simulation \cite{meyer2014simulating} is used to generate the velocity field.

Control parameters for this problem are the turbulent Mach number,
\begin{align}
 {\rm Ma}_t = \frac{\sqrt{\langle \bm{u}\cdot\bm{u}\rangle}}{\langle c_s \rangle},   
\end{align}
and the Reynolds number based on the Taylor microscale,
  \begin{align}
 	{\rm Re}_\lambda = \frac{\langle\rho\rangle u_{rms}\lambda}{\mu_0}, \label{eq:Taylor_Re}
 \end{align}
where $u_{rms} = \sqrt{{\langle \bm{u}\cdot\bm{u}\rangle}/{3}}$ is the root mean square (rms) of the velocity and notation $\langle\dots\rangle$ stands for the volume averaging over the entire computational domain while $\lambda$ is the Taylor microscale,
 \begin{align}
   \lambda = \frac{\langle u_x^2\rangle}{\langle(\partial_x u_x)^2\rangle}.
 \end{align}
The dynamic viscosity is following a power law dependence on temperature,
\begin{align}
    \mu = \mu_0\left(\frac{T}{T_0}\right)^{0.76},
\end{align}
with $T_0$ being the initial temperature.
The Prandtl number for all the simulations is ${\rm Pr} = 0.7$ in accordance with the DNS\cite{samtaney2001direct}.

\subsubsection{Low turbulent Mach number}
The simulation is first performed at a relatively low turbulent Mach number ${\rm Ma}_t=0.3$ with ${\rm Re}_\lambda=72$, $\kappa_0=8({2\pi/}{L})$ and initial temperature $T_0=0.15$. 
\begin{figure}[]
		\centering
		\includegraphics[width=0.5\textwidth]{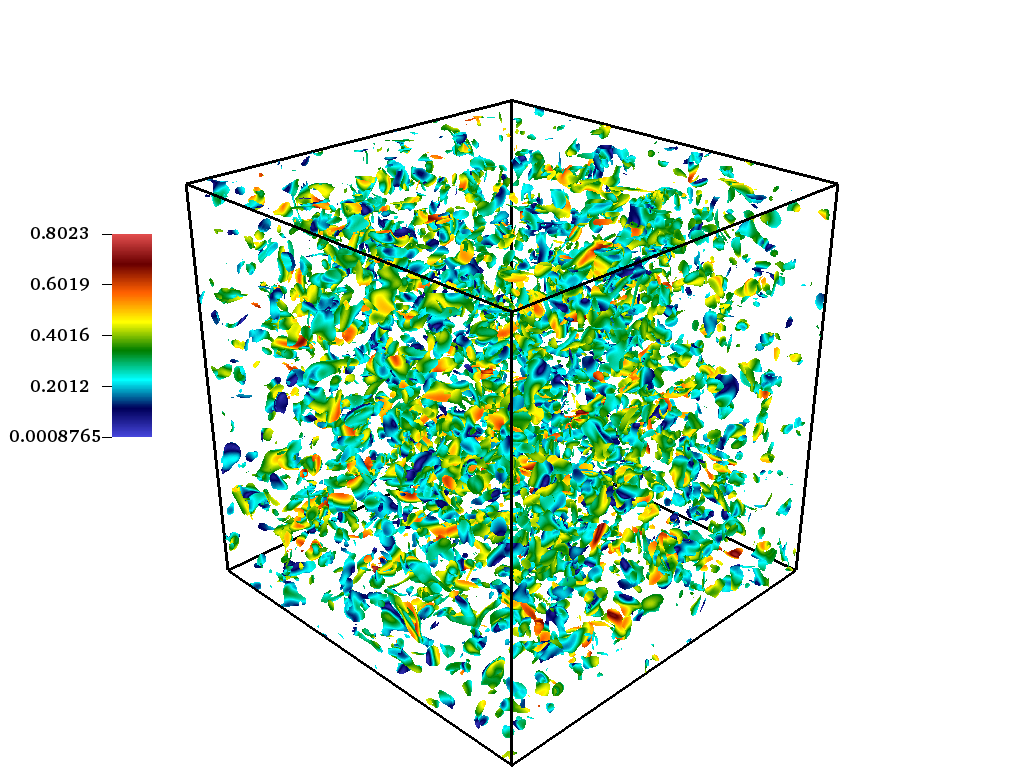}
		\caption{Iso-surface of velocity divergence $\bgrad\cdot\bm{u}=0.005$, colored by local Mach number for compressible decaying turbulence at ${\rm Ma}_t=0.3$, 
			${\rm Re}_\lambda=72$ and $t^*=0.4$.}
		\label{fig:HIT_03}
\end{figure}
Fig.\ \ref{fig:HIT_03} illustrates the instantaneous iso-surface of the velocity divergence $\bgrad\cdot\bm{u}$ colored with the local Mach number at the non-dimensional time $t^* = t/\tau=0.4$, where $\tau=L_I/u_{rms,0}$ is the large eddy turnover time defined based on the initial rms of the velocity and the integral length scale,
 \begin{align}
     L_I = \frac{3\pi\int_0^\infty \left[E(\kappa)/\kappa\right] d\kappa}{4\int_0^\infty E(\kappa)d\kappa} = \frac{\sqrt{2\pi}}{\kappa_0}.
 \end{align}
It is observed that in this case, the flow is in a moderately compressible to a high-subsonic regime, with the maximum local Mach number ${\rm Ma}_{max}\sim 0.8$.
\begin{figure}[h]
		\centering
		\includegraphics[width=0.5\textwidth]{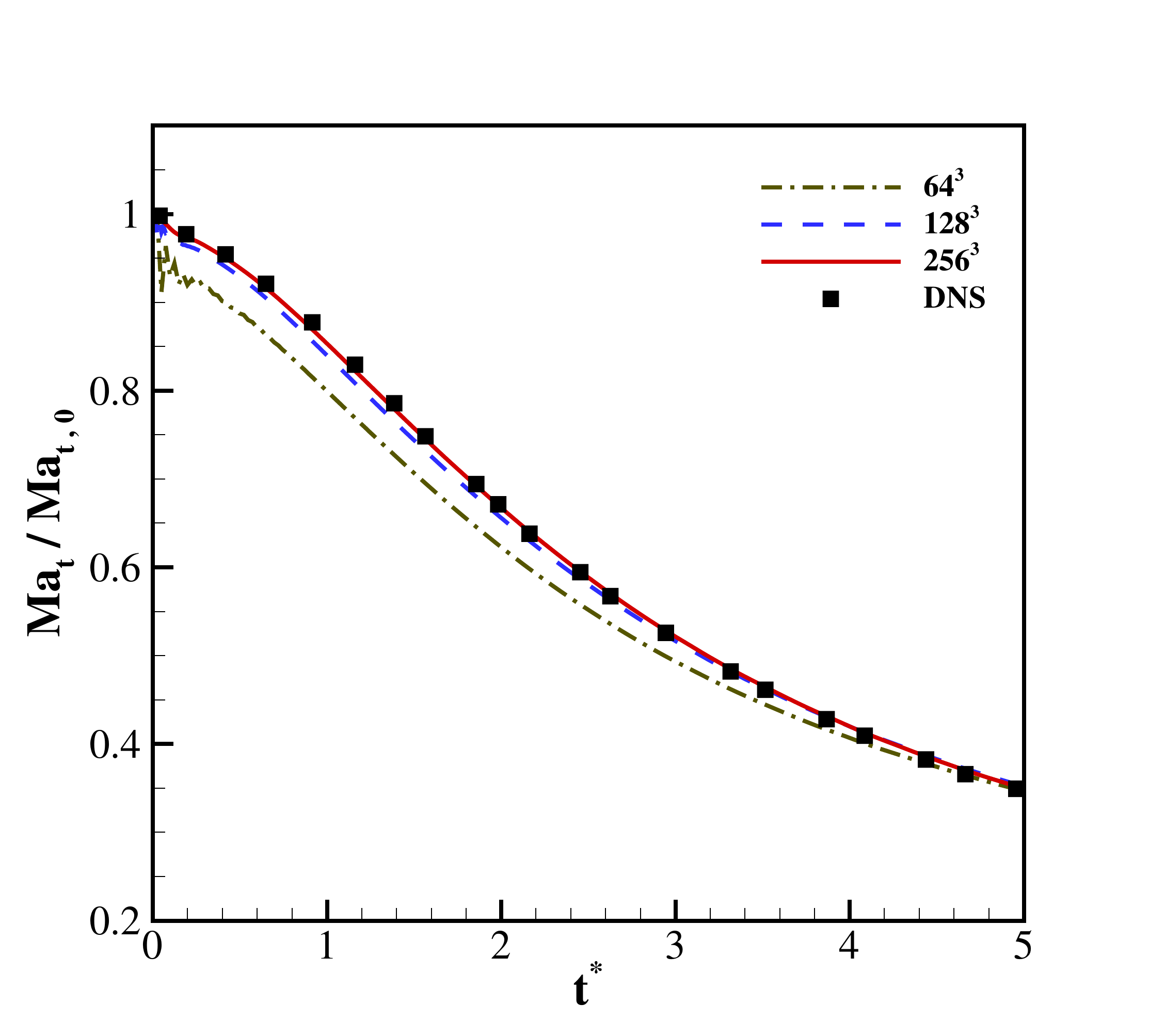}
		\caption{Decay of the turbulent Mach number for compressible decaying turbulence at ${\rm Ma}_t=0.3$ and ${\rm Re}_\lambda=72$. Lines: present model; symbol: DNS \cite{samtaney2001direct}. }
		\label{fig:Ma_03}
\end{figure}
\begin{figure}[]
		\centering
		\includegraphics[width=0.5\textwidth]{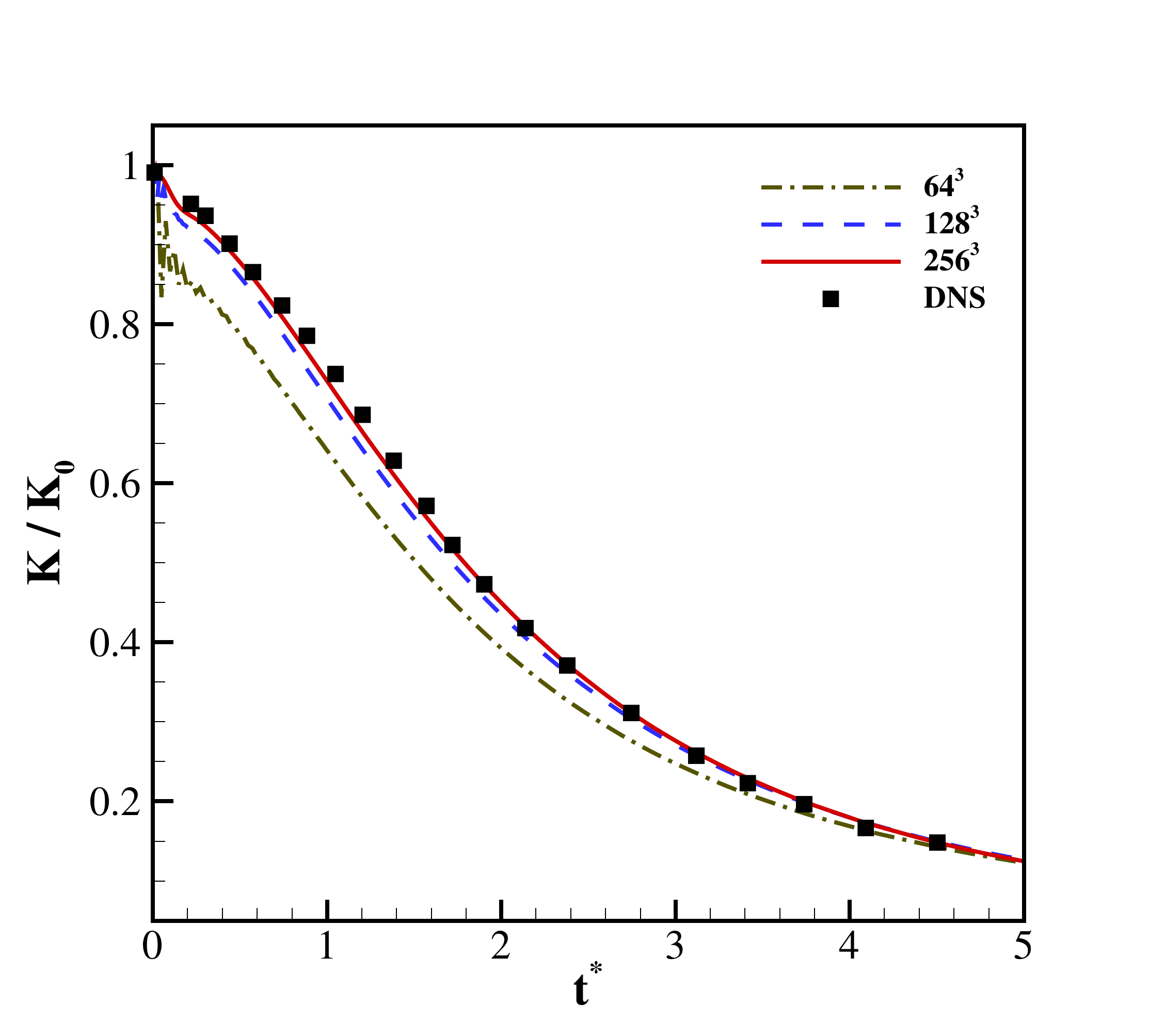}
		\caption{Decay of the turbulent kinetic energy for compressible decaying turbulence at ${\rm Ma}_t=0.3$ and ${\rm Re}_\lambda=72$. Lines: present model; symbol: DNS \cite{samtaney2001direct}. }
		\label{fig:KE_03}
\end{figure}

In order to quantify the validity of the model, a grid convergence study is performed by using three domain sizes, $64^3$, $128^3$ and $256^3$. The decay of the turbulent Mach number and of the turbulent kinetic energy $K = 1/2 \langle \rho u^2\rangle$ are shown in Fig.\ \ref{fig:Ma_03} and Fig.\ \ref{fig:KE_03}, where the convergence to the DNS results \cite{samtaney2001direct} can be observed. 
\begin{figure}[h]
		\centering
		\includegraphics[width=0.5\textwidth]{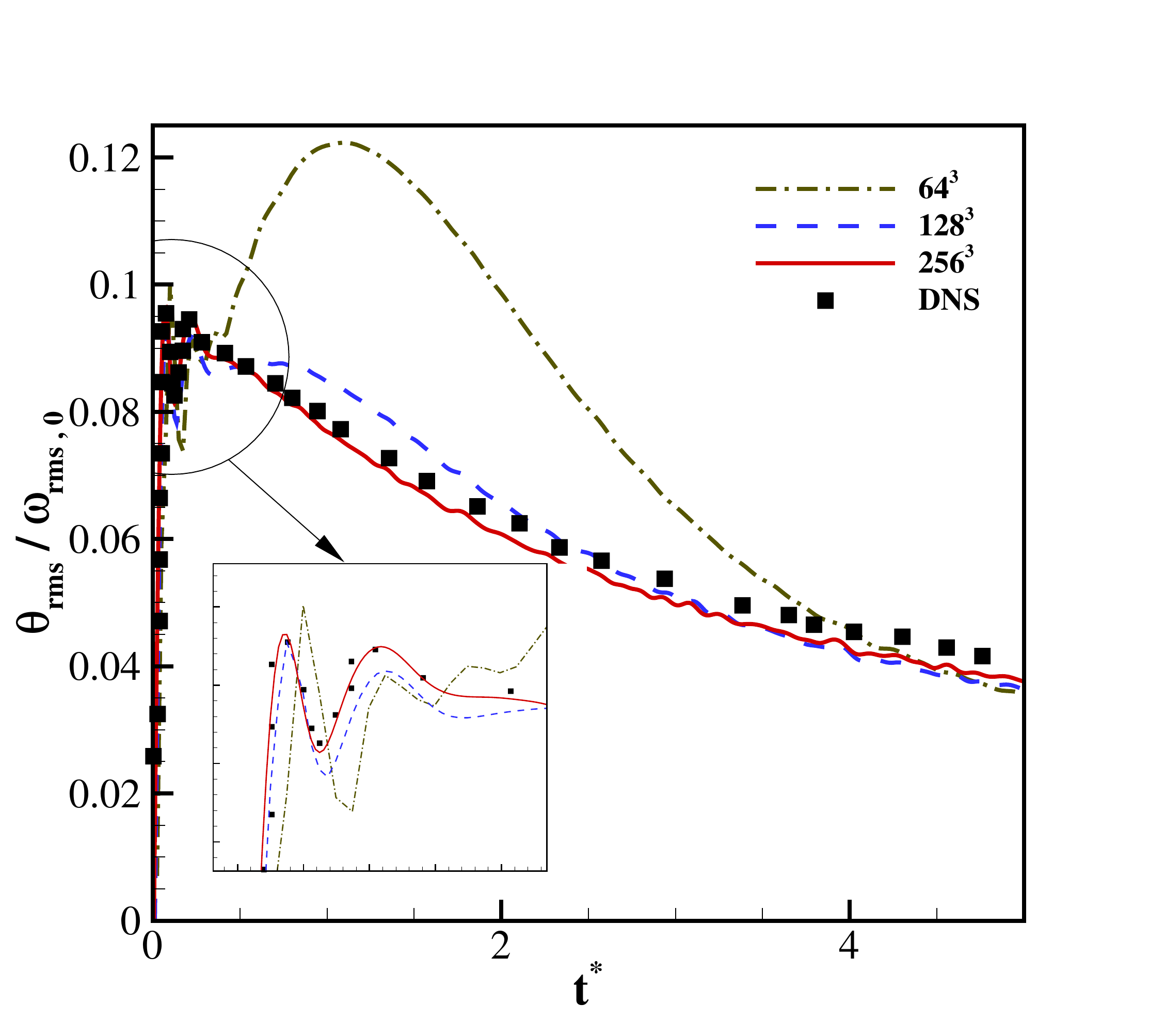}
		\caption{Time history of root mean square of dilitation for compressible decaying turbulence at ${\rm Ma}_t=0.3$ and ${\rm Re}_\lambda=72$. Lines: present model; symbol: DNS \cite{samtaney2001direct}.}
		\label{fig:Dil_03}
\end{figure}

\begin{figure}[h]
		\centering
		\includegraphics[width=0.5\textwidth]{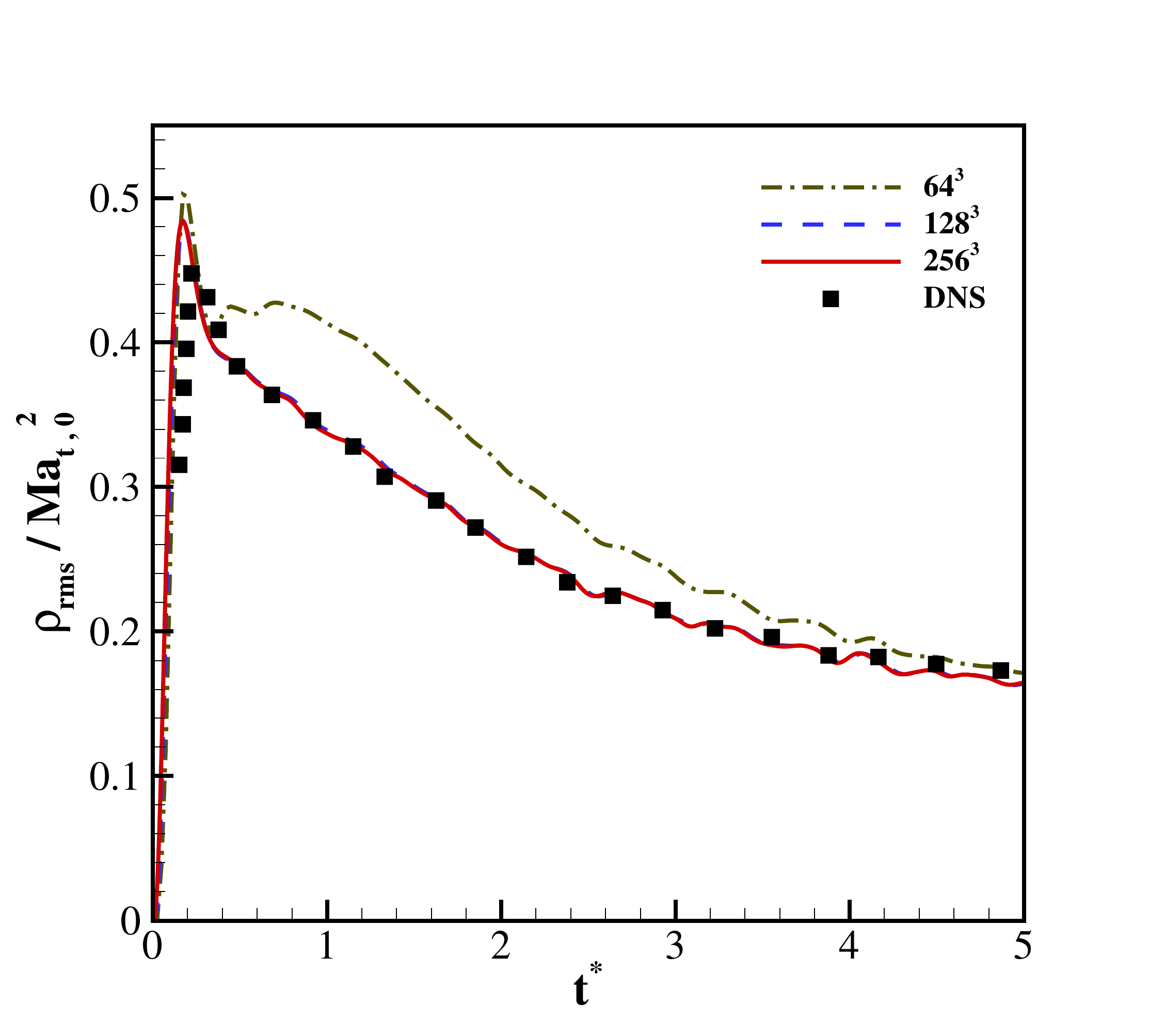}
		\caption{Time history of root mean square of density for compressible decaying turbulence at ${\rm Ma}_t=0.3$ and ${\rm Re}_\lambda=72$. Lines: present model; symbol: DNS \cite{samtaney2001direct}.}
		\label{fig:Rho_03}
\end{figure}

To assess the effect of compressibilty, time evolution of the rms of dilatation,
\begin{align}
    \theta_{rms} = \sqrt{\langle (\bgrad\cdot\bm{u})^2\rangle},
\end{align}
is compared in Fig. \ref{fig:Dil_03} with the DNS, where dilatation is normalized with the initial rms of vorticity, 
$\omega_{rms,0} = \sqrt{\langle |\bm{\omega}_0|^2\rangle}$, and $\bm{\omega} = \nabla \times \bm{u}$. Strong compressibility effects can be seen at the initial stage, where dilatation rapidly increases  from its initial value $\theta_{rms,0}=0$, followed by a monotonic decay. Furthermore, the rms of the density $\rho_{rms} = \sqrt{\langle\rho^2\rangle - \langle\rho\rangle^2}$ normalized by ${\rm Ma}_{t,0}^2$ is shown in 
Fig. \ref{fig:Rho_03}. Also here the agreement with the DNS is quite satisfactory with $256^3$ grid points.   

\begin{figure}[h]
		\centering
		\includegraphics[width=0.5\textwidth]{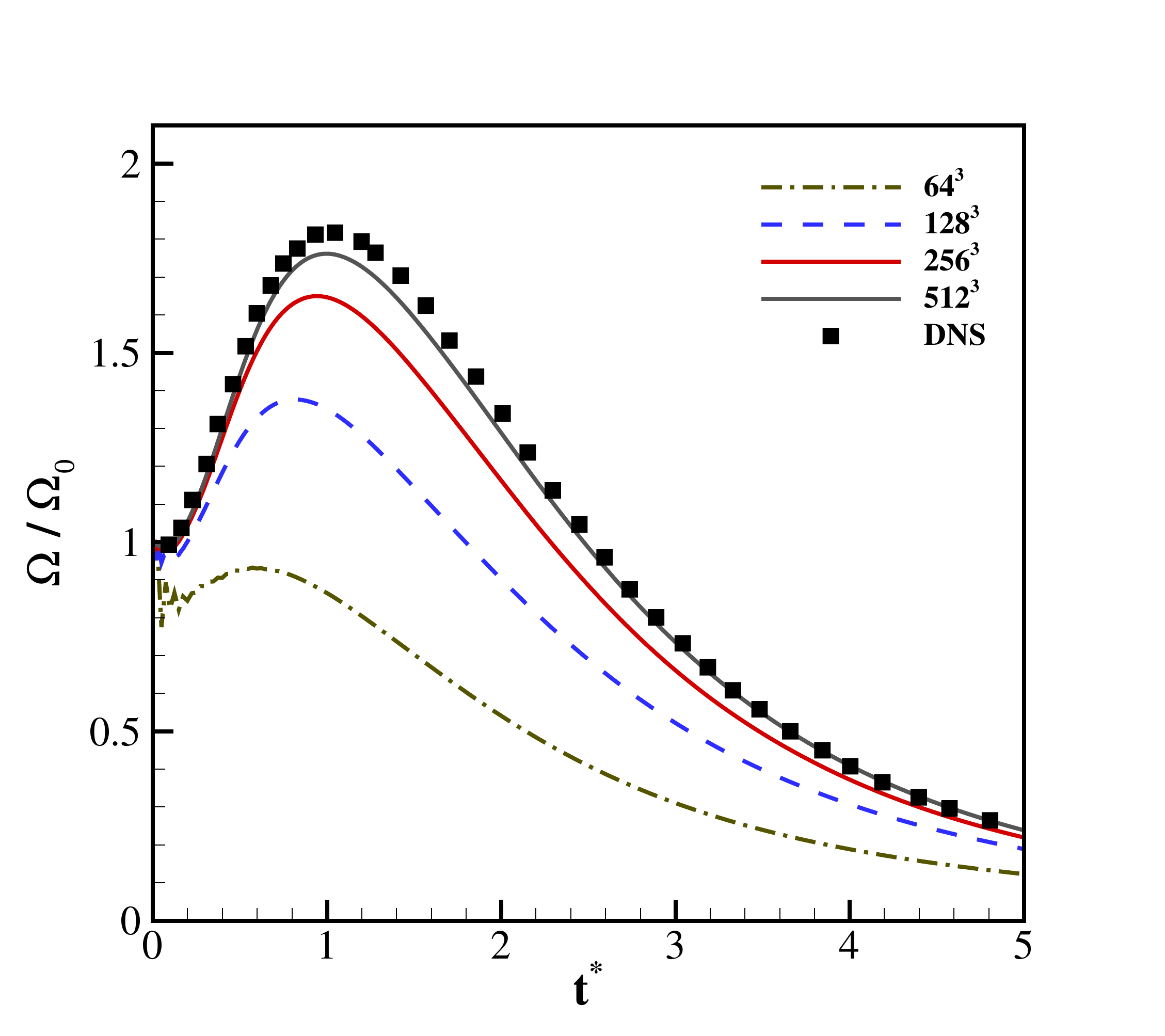}
		\caption{Time history of enstrophy for compressible decaying turbulence at ${\rm Ma}_t=0.3$ and ${\rm Re}_\lambda=72$. Lines: present model; symbol: DNS \cite{fang2014investigation}.}
		\label{fig:Diss_03}
\end{figure}

The enstrophy defined as,
\begin{align}
    \Omega = \frac{1}{2} \langle \omega^2 \rangle,
\end{align}
is a sensitive variable to analyze the performance of a numerical scheme for turbulent flows, as it is closely related to small-scale turbulence motions \cite{fang2014investigation,garnier1999use}. The temporal evolution of the enstrophy normalized with its initial value is compared in Fig. \ref{fig:Diss_03} with the DNS results of the spectral method reported in \citet{fang2014investigation}.
It can be seen that in all cases the enstrophy increases in the beginning due to vortex stretching, which generates small-scale turbulence structures. This makes the viscous dissipation stronger, which leads to a decrease of enstrophy \cite{garnier1999use}. Furthermore, coarse simulations result in under-prediction of peak value and also fast decay rate, due to strong suppression of small-scale fluctuations. Here, contrary to the previous cases, $256^3$ grid size is not enough to accurately capture the statistics. By increasing the resolution to $512^3$, the peak value and decay rate of enstrophy can be captured with good accuracy. This further confirms the accuracy of the present model in capturing the physics of compressible turbulence.
\begin{figure}[h]
		\centering
		\includegraphics[width=0.5\textwidth]{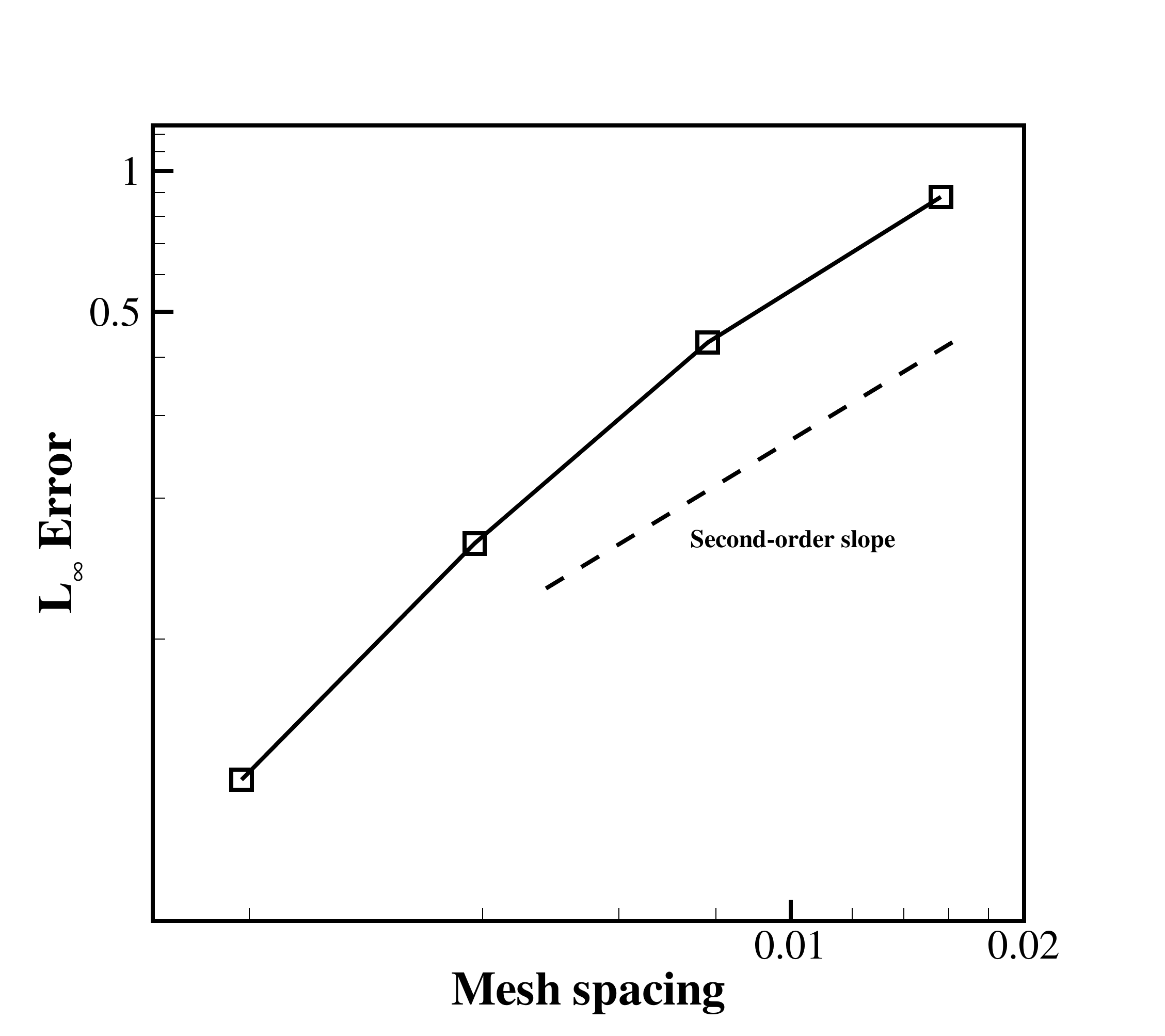}
		\caption{Convergence rate of enstrophy for grid resolutions $64^3$ to $512^3$. Symbols: $L_\infty$ error of enstrophy with respect to the DNS results; dashed line: second-order slope.}
		\label{fig:Covergence}
\end{figure}
Moreover, the convergence order of the model is evaluated based on the $L_\infty$ error of enstrophy with respect to the DNS results. As shown in Fig. \ref{fig:Covergence}, the overall accuracy in space is slightly below second-order.  

\subsubsection{Effect of deviation discretization on the accuracy}
As pointed out earlier, first-order upwind discretization of the deviation term is necessary for preventing the Gibbs phenomenon and maintaining the stability of the model in supersonic regime. %

Here, we investigate the effect of the discretization scheme on the accuracy in subsonic turbulent regime, by comparing the results to the case with second-order central evaluation of derivatives of deviation term.
\begin{figure}[h]
		\centering
		\includegraphics[width=0.5\textwidth]{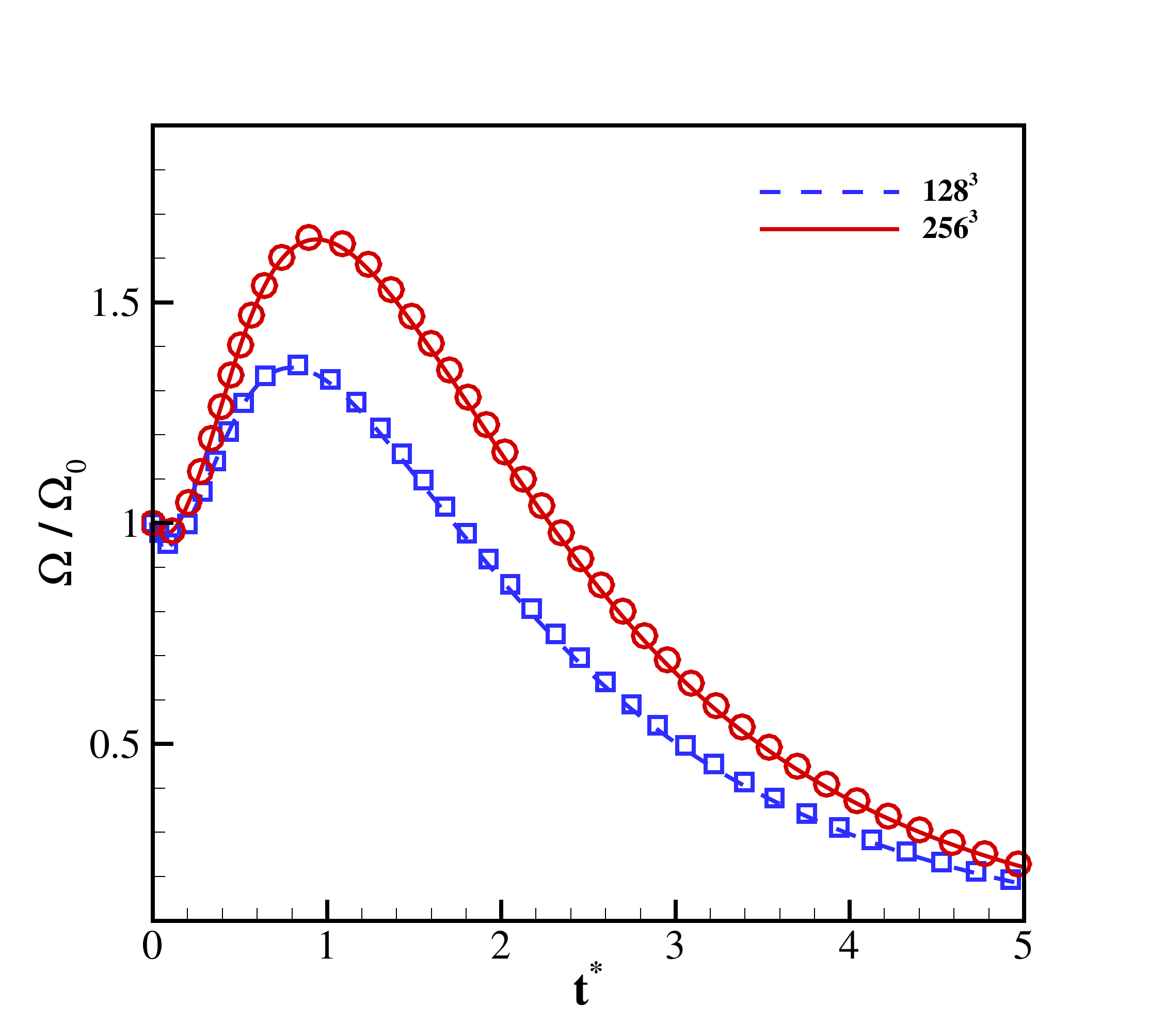}
		\caption{Effect of deviation discretization on the enstrophy evolution for compressible decaying turbulence at ${\rm Ma}_t=0.3$ and ${\rm Re}_\lambda=72$. Lines: present model with first-order upwind discretization of deviation term; symbols: present model with second-order central difference discretization of deviation term.}
		\label{fig:Diss_03_central}
\end{figure}
It can be seen from Fig.\ \ref{fig:Diss_03_central} that, the time history of enstrophy is insensitive to the evaluation of deviation term. All other turbulence statistics showed similar behaviour, but are not presented here for the sake of brevity. This indicates that the use of first-order scheme does not degrade the formal accuracy of the solver (shown in Fig.\ \ref{fig:Covergence}), although it provides sufficient dissipation for stabilizing the solver and capturing the shock.

\subsubsection{High turbulent Mach number}
We now move on to a higher turbulent Mach number.
It is well known that at sufficient high turbulent Mach numbers, random shock waves commonly known as eddy-shocklets appear in the flow \cite{lee1991eddy,samtaney2001direct,johnsen2010assessment}, due to compressiblity and turbulent motions. This scenario can, therefore, be considered as a rigorous test case for the validity of the present model.

We increase the turbulent Mach number to ${\rm Ma}_t = 0.5$ and perform the simulation with $256^3$ and $512^3$ grid points and the same Reynolds number ${\rm Re}_\lambda = 72$. %
\begin{figure}[h]
		\centering
		\includegraphics[width=0.5\textwidth]{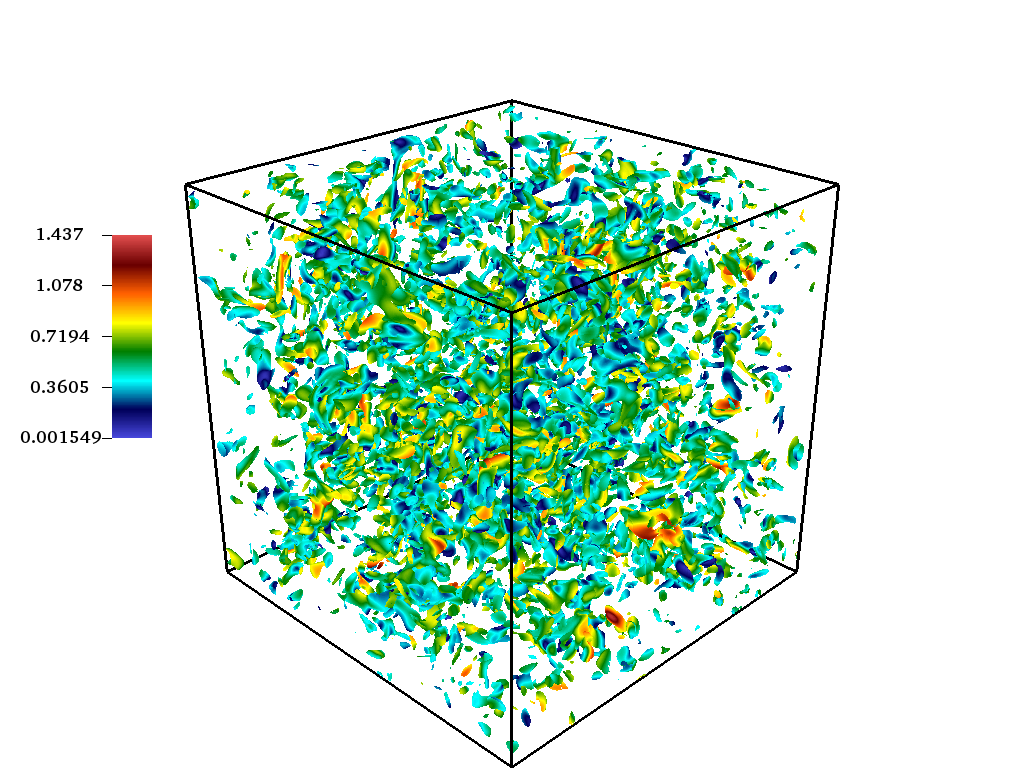}
		\caption{Iso-surface of velocity divergence $\bgrad\cdot\bm{u}=0.015$, colored by local Mach number for compressible decaying turbulence at ${\rm Ma}_t=0.4$, ${\rm Re}_\lambda=72$ and $t^*=0.5$.}
		\label{fig:HIT_05}
\end{figure}
The iso-surface of the velocity divergence colored by local Mach number is shown in Fig.\ \ref{fig:HIT_05}, which confirms the presence of local supersonic regions during the decay. 
\begin{figure}[h]
		\centering
		\includegraphics[width=0.5\textwidth]{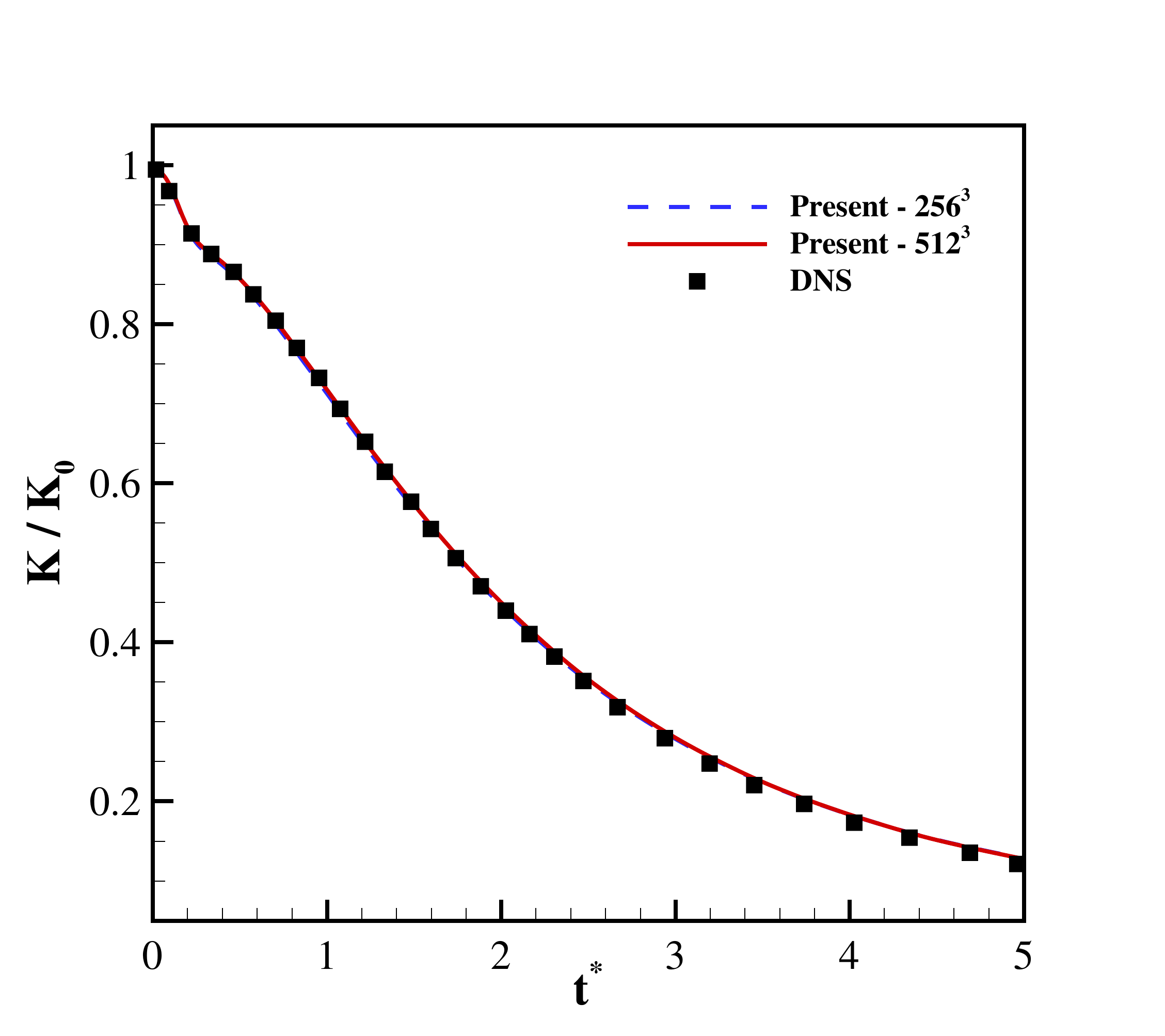}
		\caption{Decay of the turbulent kinetic energy for compressible decaying turbulence at ${\rm Ma}_t=0.5$ and ${\rm Re}_\lambda=72$. Lines: present model; symbol: DNS \cite{samtaney2001direct}.}
		\label{fig:KE_05}
\end{figure}
\begin{figure}[h]
		\centering
		\includegraphics[width=0.5\textwidth]{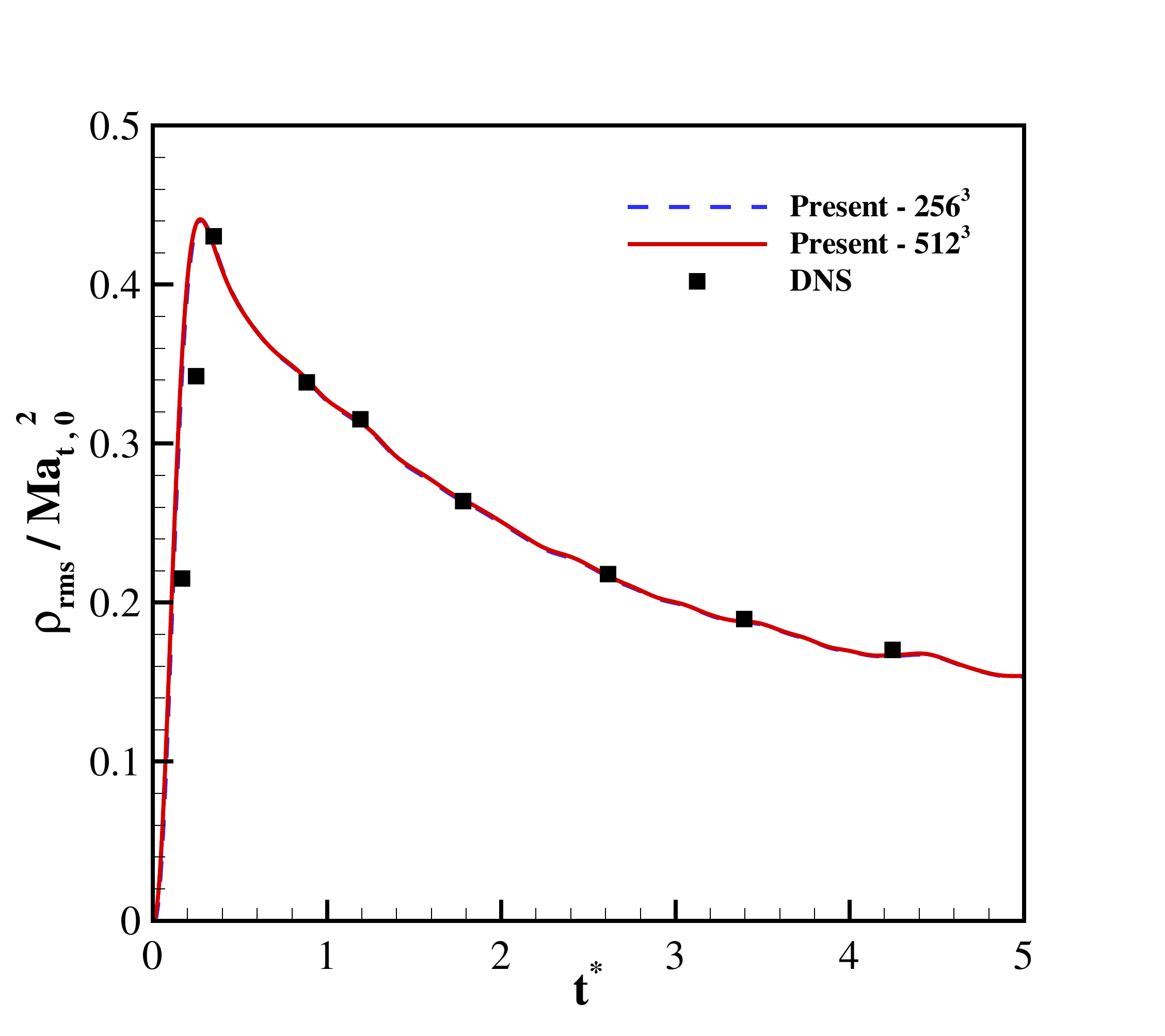}
		\caption{Time history of root mean square of density for compressible decaying turbulence at ${\rm Ma}_t=0.5$ and ${\rm Re}_\lambda=72$. Lines: present model; symbol: DNS \cite{samtaney2001direct}.}
		\label{fig:Rho_05}
\end{figure}
\begin{figure}[h]
		\centering
		\includegraphics[width=0.5\textwidth]{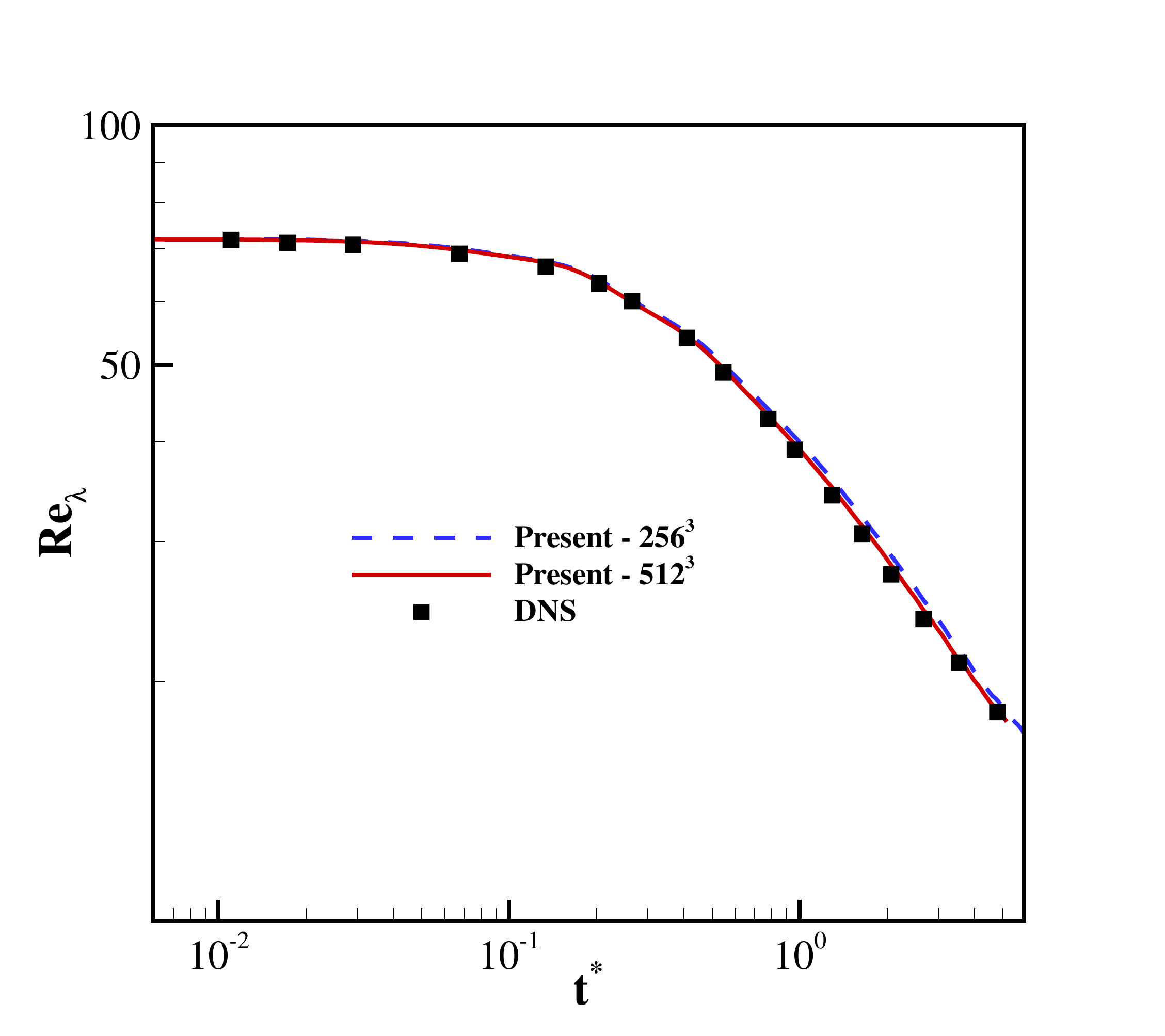}
		\caption{Time history of  Taylor microscale Reynolds number for compressible decaying turbulence at ${\rm Ma}_t=0.5$ and ${\rm Re}_\lambda=72$. Lines: present model; symbol: DNS \cite{samtaney2001direct}.}
		\label{fig:Rho_05}
\end{figure}
Moreover, to show that the model can accurately predict turbulent statistics in the presence of shocks, time evolution of the turbulent kinetic energy, rms of density and Taylor microscale Reynolds number are plotted in Figs. \ref{fig:KE_05}, \ref{fig:Rho_05} and \ref{}, respectively. Here also the results agree well with the reference DNS results. 

		\label{fig:Spectrum_05}

As a final validation case, we investigate the performance of the model at a relatively high Reynolds number of $\rm Re_\lambda=175$, while keeping the turbulent Mach number high enough $\rm Ma_t=0.488$. The initial spectrum peaks at $\kappa_0 = 4$ in this case. 
\begin{figure}[h]
		\centering
		\includegraphics[width=0.5\textwidth]{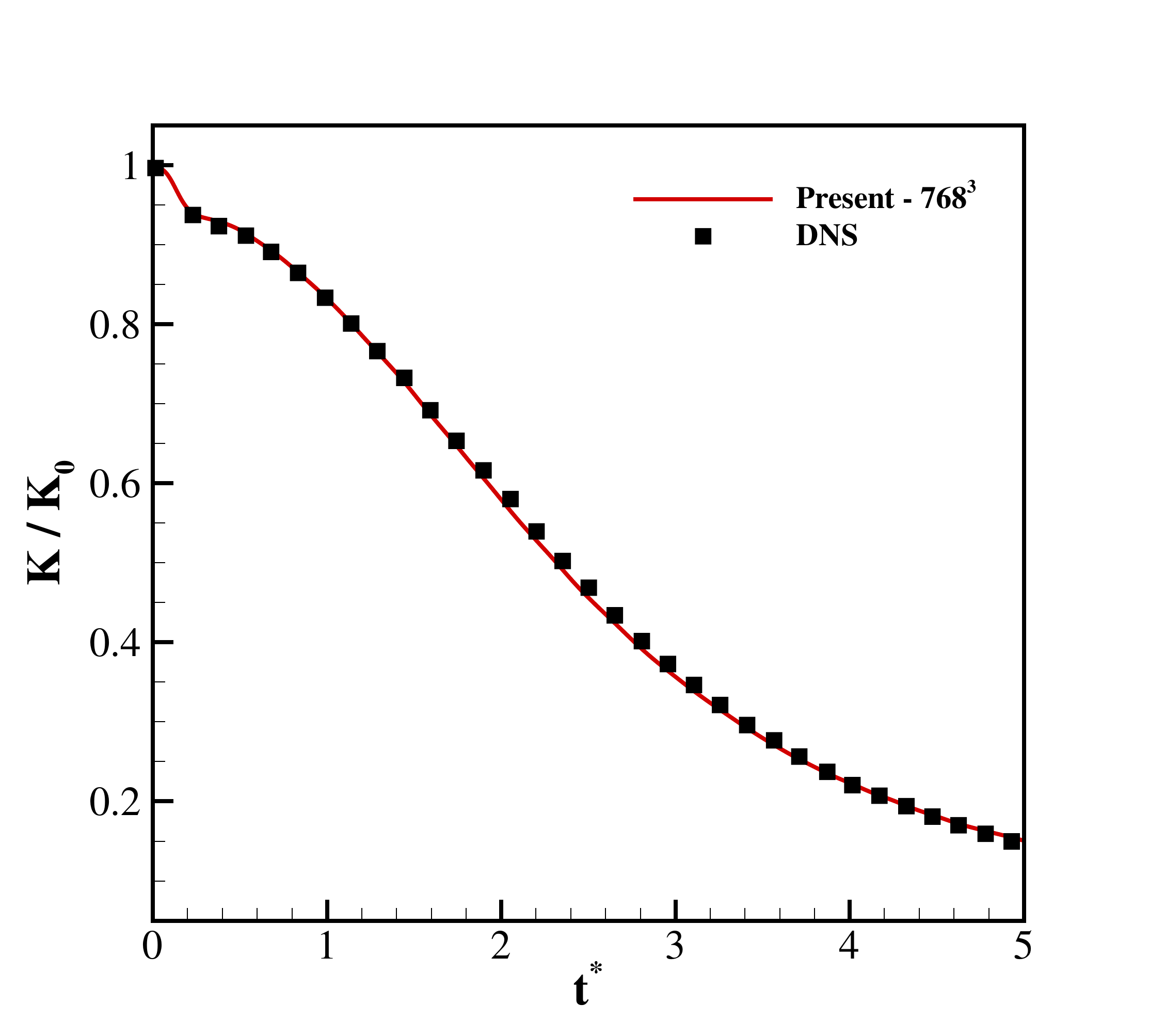}
		\caption{Decay of the turbulent kinetic energy for compressible decaying turbulence at ${\rm Ma}_t=0.488$ and ${\rm Re}_\lambda=175$. Line: present model; symbol: DNS \cite{samtaney2001direct}.}
		\label{fig:KE_0488}
\end{figure}
\begin{figure}[h]
		\centering
		\includegraphics[width=0.5\textwidth]{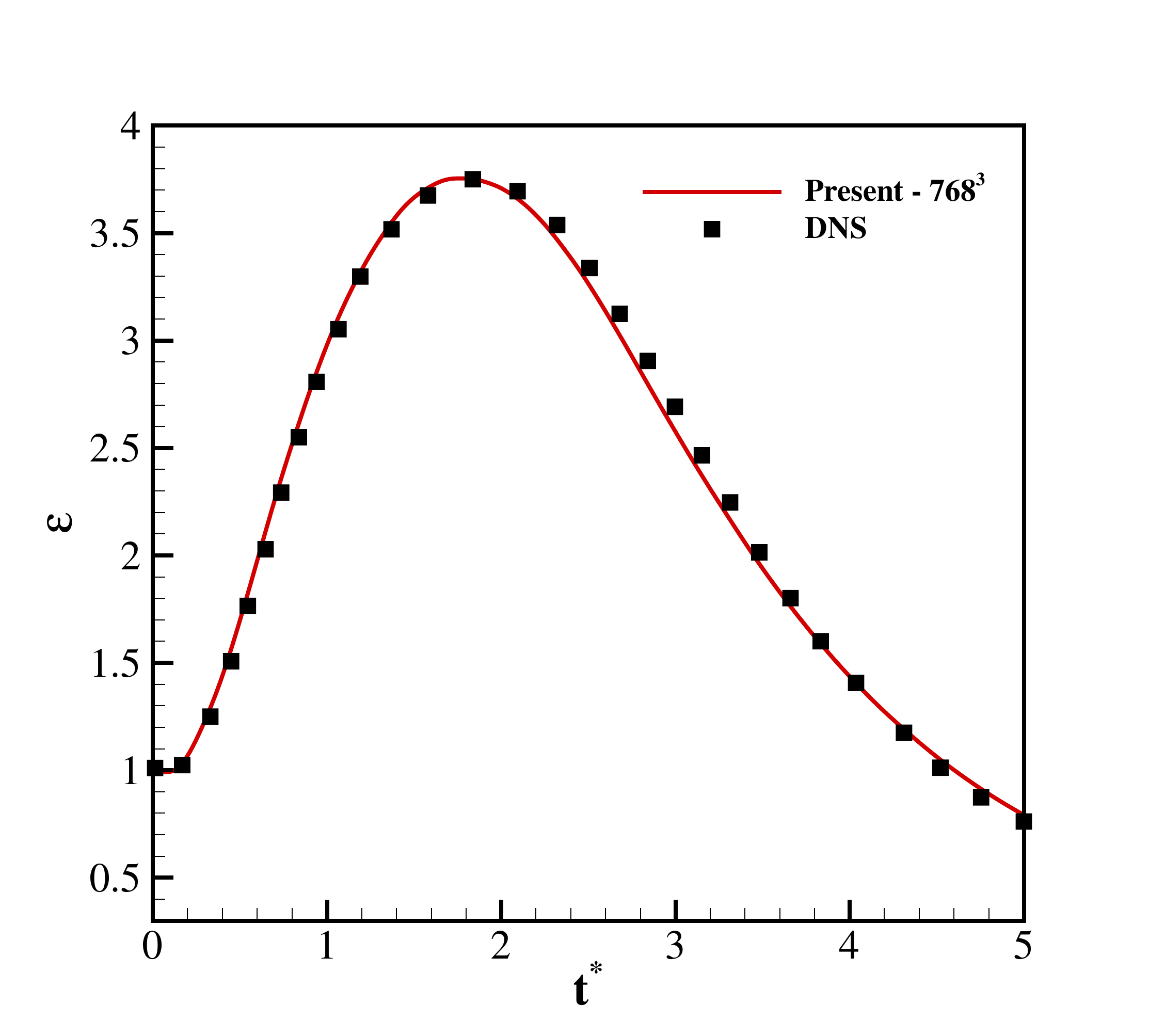}
		\caption{Time history of the dissipation rate for compressible decaying turbulence at ${\rm Ma}_t=0.488$ and ${\rm Re}_\lambda=175$. Line: present model; symbol: DNS \cite{samtaney2001direct}.}
		\label{fig:Diss_0488}
\end{figure}
\begin{figure}[h]
		\centering
		\includegraphics[width=0.5\textwidth]{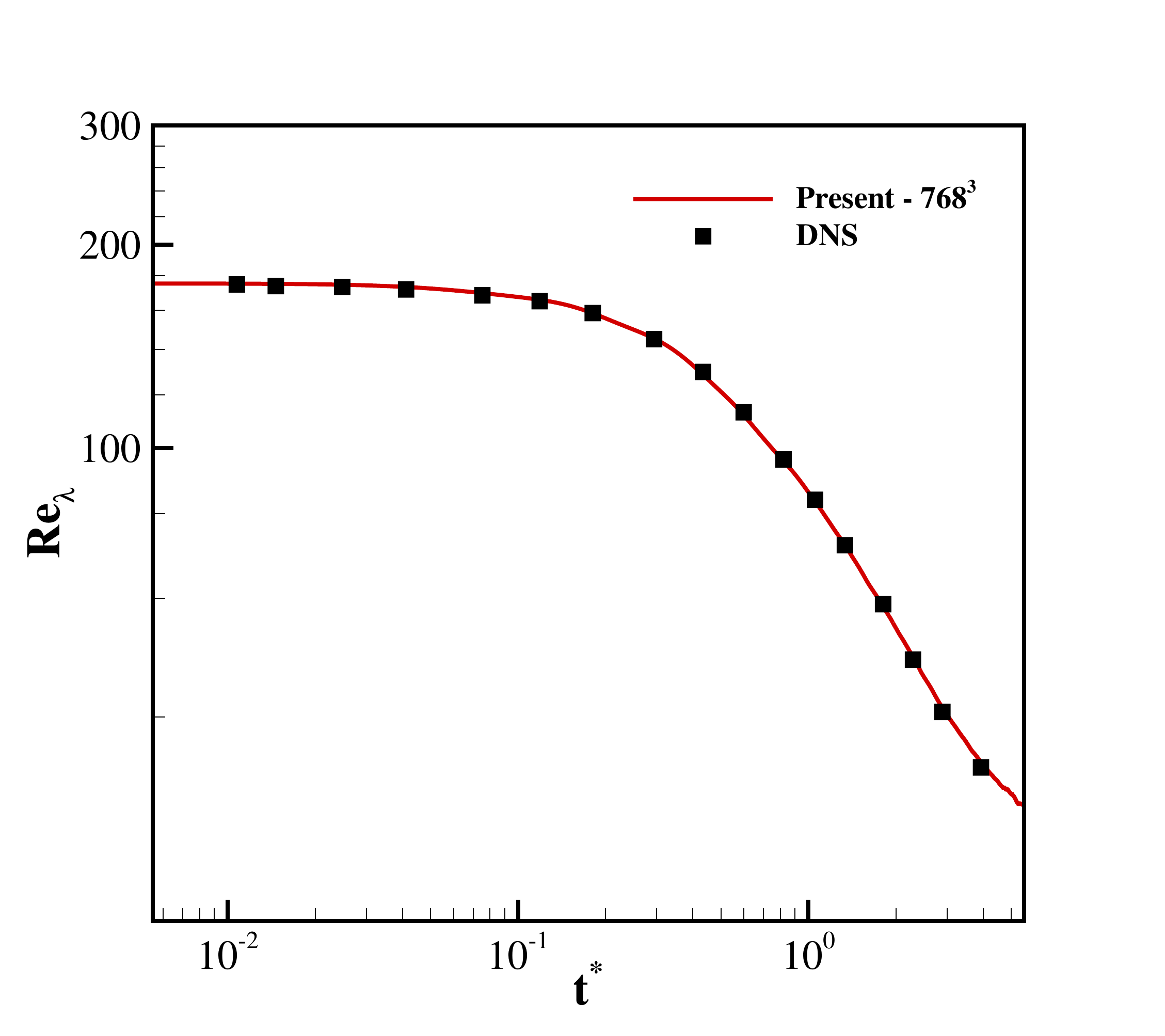}
		\caption{Time history of  Taylor microscale Reynolds number for compressible decaying turbulence at ${\rm Ma}_t=0.488$ and ${\rm Re}_\lambda=175$. Line: present model; symbol: DNS \cite{samtaney2001direct}.}
		\label{fig:Re_0488}
\end{figure}

\begin{figure}[h]
		\centering
		\includegraphics[width=0.5\textwidth]{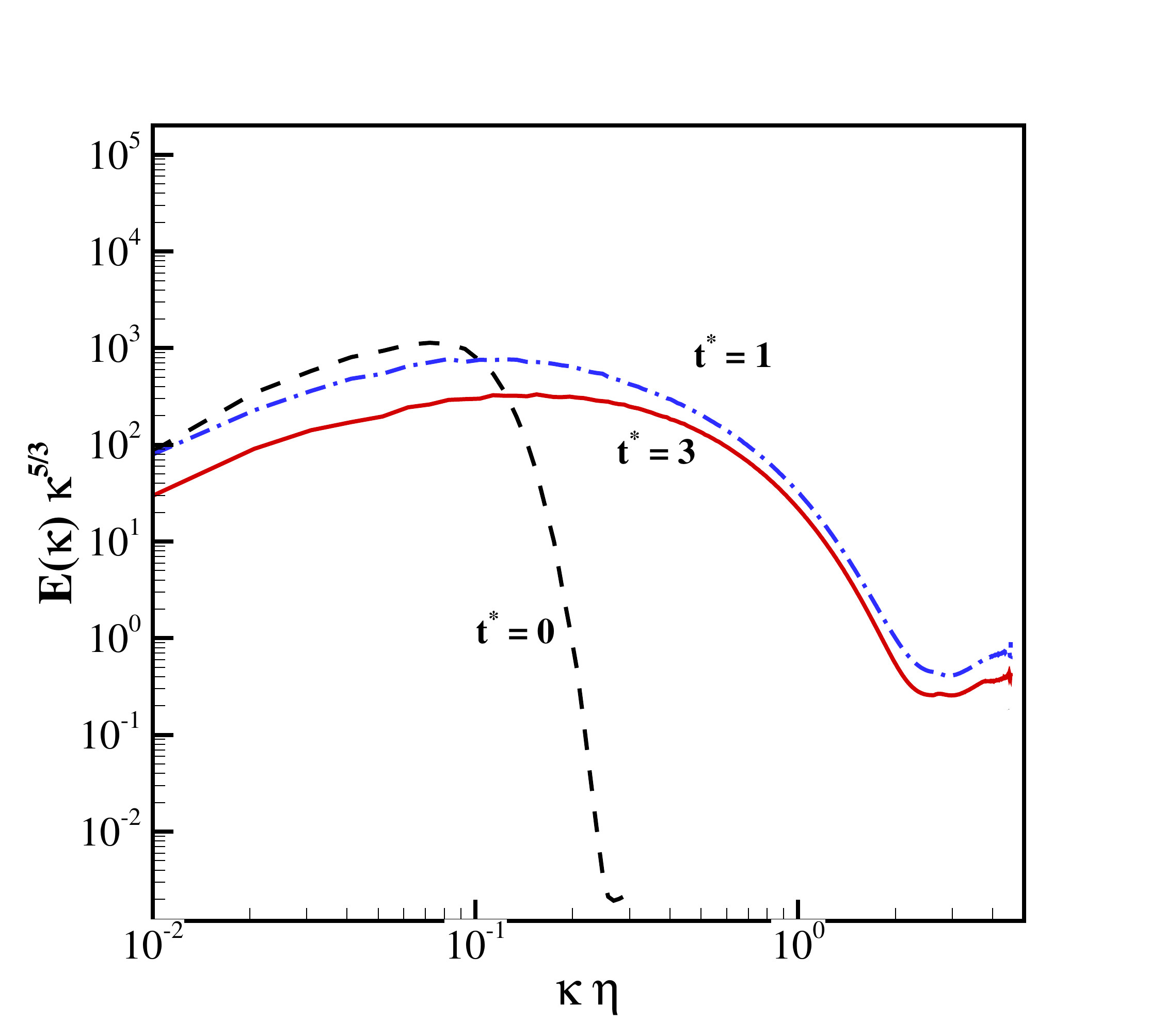}
		\caption{Energy spectrum at various times ($t^*=0, 1$ and $3$) for compressible decaying turbulence at $Ma_t=0.488$ and $Re_\lambda=175$. Dashed line is the initial spectrum ($t^*=0$). Here, $\eta$ is the Kolmogorov length scale.}
		\label{fig:Spectrum_0488}
\end{figure}

History of turbulent kinetic energy, solenoidal dissipation rate $\epsilon = \langle \mu \omega^2 \rangle$ and Taylor miroscale Reynolds number (\ref{eq:Taylor_Re}) are plotted in Figs.\ \ref{fig:KE_0488}, \ref{fig:Diss_0488} and \ref{fig:Re_0488}, using $768^3$ grid points. The results agree well with the reference DNS solution \cite{samtaney2001direct}. The energy spectrum at various times is shown in Fig.\ref{fig:Spectrum_0488}.
It is observed that initially, large scales contain most of the energy and as time evolves the energy is transferred to small scales. Moreover, since the Reynolds number is high enough, the spectrum shows the inertial range with slope of $\kappa^{-5/3}$ which further confirms the accuracy of the results and shows the ability of the model in capturing broadband turbulent motions in the presence of shocks.

\section{Conclusion} \label{Sec:Conclusion}
In this work, we proposed a lattice Boltzmann framework for the simulation of compressible flows on standard lattices. The product-form factorization was used to represent all pertinent equilibrium and quasi-equilibrium populations. The well-known anomaly of the standard lattices was eliminated by redefining the diagonal components of the equilibrium pressure tensor through adding an appropriate correction term. The analysis of the model was conducted through simulation of the Sod's shock tube, sound generation in shock-vortex interaction and compressible decaying homogeneous isotropic turbulence. 

It was demonstrated that the present fully on-lattice model with the single relaxation time LBGK collision term is able to properly predict the relevant features of the compressible flows. In particular, the model can capture moderately supersonic shock waves up to ${\rm Ma} \sim 1.5$.
Furthermore, the simulation of compressible decaying turbulence
demonstrated that the model can accurately capture compressibilty effects, turbulence fluctuations and shocks. It was shown that the model performs well even at high turbulent Mach number, where eddy-shocklets exist in the flow field and interact with turbulence. The results of the model were found to be in good agreement with DNS results.

Overall, the promising results of the proposed model on standard lattices open interesting prospects towards the numerical simulation of more complex applications such as compressible jet flow \cite{vuorinen2013large} or shock boundary-layer interaction \cite{pirozzoli2010direct}. Moreover, the model could be augmented with more sophisticated collision terms in order to enhance the stability of under-resolved simulations. This will be the focus of our future research.

\begin{acknowledgments}
This work was supported by the ETH research grant ETH-13 17-1 and the European Research Council (ERC) Advanced Grant No. 834763-PonD. The computational resources at the Swiss National Super Computing Center CSCS were provided under the grant s897.\\
\end{acknowledgments}

\bibliography{Bib}%

\begin{thebibliography}{48}%
\makeatletter
\providecommand \@ifxundefined [1]{%
 \@ifx{#1\undefined}
}%
\providecommand \@ifnum [1]{%
 \ifnum #1\expandafter \@firstoftwo
 \else \expandafter \@secondoftwo
 \fi
}%
\providecommand \@ifx [1]{%
 \ifx #1\expandafter \@firstoftwo
 \else \expandafter \@secondoftwo
 \fi
}%
\providecommand \natexlab [1]{#1}%
\providecommand \enquote  [1]{``#1''}%
\providecommand \bibnamefont  [1]{#1}%
\providecommand \bibfnamefont [1]{#1}%
\providecommand \citenamefont [1]{#1}%
\providecommand \href@noop [0]{\@secondoftwo}%
\providecommand \href [0]{\begingroup \@sanitize@url \@href}%
\providecommand \@href[1]{\@@startlink{#1}\@@href}%
\providecommand \@@href[1]{\endgroup#1\@@endlink}%
\providecommand \@sanitize@url [0]{\catcode `\\12\catcode `\$12\catcode
  `\&12\catcode `\#12\catcode `\^12\catcode `\_12\catcode `\%12\relax}%
\providecommand \@@startlink[1]{}%
\providecommand \@@endlink[0]{}%
\providecommand \url  [0]{\begingroup\@sanitize@url \@url }%
\providecommand \@url [1]{\endgroup\@href {#1}{\urlprefix }}%
\providecommand \urlprefix  [0]{URL }%
\providecommand \Eprint [0]{\href }%
\providecommand \doibase [0]{https://doi.org/}%
\providecommand \selectlanguage [0]{\@gobble}%
\providecommand \bibinfo  [0]{\@secondoftwo}%
\providecommand \bibfield  [0]{\@secondoftwo}%
\providecommand \translation [1]{[#1]}%
\providecommand \BibitemOpen [0]{}%
\providecommand \bibitemStop [0]{}%
\providecommand \bibitemNoStop [0]{.\EOS\space}%
\providecommand \EOS [0]{\spacefactor3000\relax}%
\providecommand \BibitemShut  [1]{\csname bibitem#1\endcsname}%
\let\auto@bib@innerbib\@empty
\bibitem [{\citenamefont {Caughey}(2003)}]{caughey2003computational}%
  \BibitemOpen
  \bibfield  {author} {\bibinfo {author} {\bibfnamefont {D.~A.}\ \bibnamefont
  {Caughey}},\ }\href@noop {} {\emph {\bibinfo {title} {Computational
  aerodynamics}}}\ (\bibinfo  {publisher} {Elsevier},\ \bibinfo {year}
  {2003})\BibitemShut {NoStop}%
\bibitem [{\citenamefont {von Neumann}\ and\ \citenamefont
  {Richtmyer}(1950)}]{vonneumann1950method}%
  \BibitemOpen
  \bibfield  {author} {\bibinfo {author} {\bibfnamefont {J.}~\bibnamefont {von
  Neumann}}\ and\ \bibinfo {author} {\bibfnamefont {R.~D.}\ \bibnamefont
  {Richtmyer}},\ }\bibfield  {title} {\enquote {\bibinfo {title} {A method for
  the numerical calculation of hydrodynamic shocks},}\ }\href@noop {}
  {\bibfield  {journal} {\bibinfo  {journal} {Journal of applied physics}\
  }\textbf {\bibinfo {volume} {21}},\ \bibinfo {pages} {232--237} (\bibinfo
  {year} {1950})}\BibitemShut {NoStop}%
\bibitem [{\citenamefont {Shu}(1999)}]{shu1999high}%
  \BibitemOpen
  \bibfield  {author} {\bibinfo {author} {\bibfnamefont {C.-W.}\ \bibnamefont
  {Shu}},\ }\bibfield  {title} {\enquote {\bibinfo {title} {{High order ENO and
  WENO schemes for computational fluid dynamics}},}\ }in\ \href@noop {} {\emph
  {\bibinfo {booktitle} {High-order methods for computational physics}}}\
  (\bibinfo  {publisher} {Springer},\ \bibinfo {year} {1999})\ pp.\ \bibinfo
  {pages} {439--582}\BibitemShut {NoStop}%
\bibitem [{\citenamefont {Subramaniam}, \citenamefont {Wong},\ and\
  \citenamefont {Lele}(2019)}]{subramaniam2019high}%
  \BibitemOpen
  \bibfield  {author} {\bibinfo {author} {\bibfnamefont {A.}~\bibnamefont
  {Subramaniam}}, \bibinfo {author} {\bibfnamefont {M.~L.}\ \bibnamefont
  {Wong}},\ and\ \bibinfo {author} {\bibfnamefont {S.~K.}\ \bibnamefont
  {Lele}},\ }\bibfield  {title} {\enquote {\bibinfo {title} {A high-order
  weighted compact high resolution scheme with boundary closures for
  compressible turbulent flows with shocks},}\ }\href@noop {} {\bibfield
  {journal} {\bibinfo  {journal} {Journal of Computational Physics}\ }\textbf
  {\bibinfo {volume} {397}},\ \bibinfo {pages} {108822} (\bibinfo {year}
  {2019})}\BibitemShut {NoStop}%
\bibitem [{\citenamefont {Fu}, \citenamefont {Hu},\ and\ \citenamefont
  {Adams}(2018)}]{fu2018new}%
  \BibitemOpen
  \bibfield  {author} {\bibinfo {author} {\bibfnamefont {L.}~\bibnamefont
  {Fu}}, \bibinfo {author} {\bibfnamefont {X.~Y.}\ \bibnamefont {Hu}},\ and\
  \bibinfo {author} {\bibfnamefont {N.~A.}\ \bibnamefont {Adams}},\ }\bibfield
  {title} {\enquote {\bibinfo {title} {{A new class of adaptive high-order
  targeted ENO schemes for hyperbolic conservation laws}},}\ }\href@noop {}
  {\bibfield  {journal} {\bibinfo  {journal} {Journal of Computational
  Physics}\ }\textbf {\bibinfo {volume} {374}},\ \bibinfo {pages} {724--751}
  (\bibinfo {year} {2018})}\BibitemShut {NoStop}%
\bibitem [{\citenamefont {Fu}(2019)}]{fu2019very}%
  \BibitemOpen
  \bibfield  {author} {\bibinfo {author} {\bibfnamefont {L.}~\bibnamefont
  {Fu}},\ }\bibfield  {title} {\enquote {\bibinfo {title} {{A very-high-order
  TENO scheme for all-speed gas dynamics and turbulence}},}\ }\href@noop {}
  {\bibfield  {journal} {\bibinfo  {journal} {Computer Physics Communications}\
  }\textbf {\bibinfo {volume} {244}},\ \bibinfo {pages} {117--131} (\bibinfo
  {year} {2019})}\BibitemShut {NoStop}%
\bibitem [{\citenamefont {Haga}\ and\ \citenamefont
  {Kawai}(2019)}]{haga2019robust}%
  \BibitemOpen
  \bibfield  {author} {\bibinfo {author} {\bibfnamefont {T.}~\bibnamefont
  {Haga}}\ and\ \bibinfo {author} {\bibfnamefont {S.}~\bibnamefont {Kawai}},\
  }\bibfield  {title} {\enquote {\bibinfo {title} {On a robust and accurate
  localized artificial diffusivity scheme for the high-order
  flux-reconstruction method},}\ }\href@noop {} {\bibfield  {journal} {\bibinfo
   {journal} {Journal of Computational Physics}\ }\textbf {\bibinfo {volume}
  {376}},\ \bibinfo {pages} {534--563} (\bibinfo {year} {2019})}\BibitemShut
  {NoStop}%
\bibitem [{\citenamefont {Visbal}\ and\ \citenamefont
  {Gaitonde}(2005)}]{visbal2005shock}%
  \BibitemOpen
  \bibfield  {author} {\bibinfo {author} {\bibfnamefont {M.}~\bibnamefont
  {Visbal}}\ and\ \bibinfo {author} {\bibfnamefont {D.}~\bibnamefont
  {Gaitonde}},\ }\bibfield  {title} {\enquote {\bibinfo {title} {Shock
  capturing using compact-differencing-based methods},}\ }in\ \href@noop {}
  {\emph {\bibinfo {booktitle} {43rd AIAA Aerospace Sciences Meeting and
  Exhibit}}}\ (\bibinfo {year} {2005})\ p.\ \bibinfo {pages} {1265}\BibitemShut
  {NoStop}%
\bibitem [{\citenamefont {Dorschner}\ \emph {et~al.}(2016)\citenamefont
  {Dorschner}, \citenamefont {B{\"o}sch}, \citenamefont {Chikatamarla},
  \citenamefont {Boulouchos},\ and\ \citenamefont
  {Karlin}}]{dorschner2016entropic}%
  \BibitemOpen
  \bibfield  {author} {\bibinfo {author} {\bibfnamefont {B.}~\bibnamefont
  {Dorschner}}, \bibinfo {author} {\bibfnamefont {F.}~\bibnamefont
  {B{\"o}sch}}, \bibinfo {author} {\bibfnamefont {S.~S.}\ \bibnamefont
  {Chikatamarla}}, \bibinfo {author} {\bibfnamefont {K.}~\bibnamefont
  {Boulouchos}},\ and\ \bibinfo {author} {\bibfnamefont {I.~V.}\ \bibnamefont
  {Karlin}},\ }\bibfield  {title} {\enquote {\bibinfo {title} {Entropic
  multi-relaxation time lattice boltzmann model for complex flows},}\
  }\href@noop {} {\bibfield  {journal} {\bibinfo  {journal} {Journal of Fluid
  Mechanics}\ }\textbf {\bibinfo {volume} {801}},\ \bibinfo {pages} {623}
  (\bibinfo {year} {2016})}\BibitemShut {NoStop}%
\bibitem [{\citenamefont {Wagner}(2006)}]{wagner2006thermodynamic}%
  \BibitemOpen
  \bibfield  {author} {\bibinfo {author} {\bibfnamefont {A.}~\bibnamefont
  {Wagner}},\ }\bibfield  {title} {\enquote {\bibinfo {title} {{Thermodynamic
  consistency of liquid-gas lattice Boltzmann simulations}},}\ }\href@noop {}
  {\bibfield  {journal} {\bibinfo  {journal} {Physical Review E}\ }\textbf
  {\bibinfo {volume} {74}},\ \bibinfo {pages} {056703} (\bibinfo {year}
  {2006})}\BibitemShut {NoStop}%
\bibitem [{\citenamefont {Mendoza}\ \emph {et~al.}(2010)\citenamefont
  {Mendoza}, \citenamefont {Boghosian}, \citenamefont {Herrmann},\ and\
  \citenamefont {Succi}}]{mendoza2010fast}%
  \BibitemOpen
  \bibfield  {author} {\bibinfo {author} {\bibfnamefont {M.}~\bibnamefont
  {Mendoza}}, \bibinfo {author} {\bibfnamefont {B.}~\bibnamefont {Boghosian}},
  \bibinfo {author} {\bibfnamefont {H.~J.}\ \bibnamefont {Herrmann}},\ and\
  \bibinfo {author} {\bibfnamefont {S.}~\bibnamefont {Succi}},\ }\bibfield
  {title} {\enquote {\bibinfo {title} {{Fast lattice Boltzmann solver for
  relativistic hydrodynamics}},}\ }\href@noop {} {\bibfield  {journal}
  {\bibinfo  {journal} {Physical Review Letters}\ }\textbf {\bibinfo {volume}
  {105}},\ \bibinfo {pages} {014502} (\bibinfo {year} {2010})}\BibitemShut
  {NoStop}%
\bibitem [{\citenamefont {Shan}, \citenamefont {Yuan},\ and\ \citenamefont
  {Chen}(2006)}]{shan2006kinetic}%
  \BibitemOpen
  \bibfield  {author} {\bibinfo {author} {\bibfnamefont {X.}~\bibnamefont
  {Shan}}, \bibinfo {author} {\bibfnamefont {X.-F.}\ \bibnamefont {Yuan}},\
  and\ \bibinfo {author} {\bibfnamefont {H.}~\bibnamefont {Chen}},\ }\bibfield
  {title} {\enquote {\bibinfo {title} {{Kinetic theory representation of
  hydrodynamics: a way beyond the Navier--Stokes equation}},}\ }\href@noop {}
  {\bibfield  {journal} {\bibinfo  {journal} {Journal of Fluid Mechanics}\
  }\textbf {\bibinfo {volume} {550}},\ \bibinfo {pages} {413--441} (\bibinfo
  {year} {2006})}\BibitemShut {NoStop}%
\bibitem [{\citenamefont {Chikatamarla}\ and\ \citenamefont
  {Karlin}(2009)}]{chikatamarla2009lattices}%
  \BibitemOpen
  \bibfield  {author} {\bibinfo {author} {\bibfnamefont {S.~S.}\ \bibnamefont
  {Chikatamarla}}\ and\ \bibinfo {author} {\bibfnamefont {I.~V.}\ \bibnamefont
  {Karlin}},\ }\bibfield  {title} {\enquote {\bibinfo {title} {{Lattices for
  the lattice Boltzmann method}},}\ }\href@noop {} {\bibfield  {journal}
  {\bibinfo  {journal} {Physical Review E}\ }\textbf {\bibinfo {volume} {79}},\
  \bibinfo {pages} {046701} (\bibinfo {year} {2009})}\BibitemShut {NoStop}%
\bibitem [{\citenamefont {Frapolli}, \citenamefont {Chikatamarla},\ and\
  \citenamefont {Karlin}(2016{\natexlab{a}})}]{frapolli2016entropic}%
  \BibitemOpen
  \bibfield  {author} {\bibinfo {author} {\bibfnamefont {N.}~\bibnamefont
  {Frapolli}}, \bibinfo {author} {\bibfnamefont {S.~S.}\ \bibnamefont
  {Chikatamarla}},\ and\ \bibinfo {author} {\bibfnamefont {I.~V.}\ \bibnamefont
  {Karlin}},\ }\bibfield  {title} {\enquote {\bibinfo {title} {{Entropic
  lattice Boltzmann model for gas dynamics: Theory, boundary conditions, and
  implementation}},}\ }\href@noop {} {\bibfield  {journal} {\bibinfo  {journal}
  {Physical Review E}\ }\textbf {\bibinfo {volume} {93}},\ \bibinfo {pages}
  {063302} (\bibinfo {year} {2016}{\natexlab{a}})}\BibitemShut {NoStop}%
\bibitem [{\citenamefont {Wilde}\ \emph {et~al.}(2020)\citenamefont {Wilde},
  \citenamefont {Kr{\"a}mer}, \citenamefont {Reith},\ and\ \citenamefont
  {Foysi}}]{wilde2020semi}%
  \BibitemOpen
  \bibfield  {author} {\bibinfo {author} {\bibfnamefont {D.}~\bibnamefont
  {Wilde}}, \bibinfo {author} {\bibfnamefont {A.}~\bibnamefont {Kr{\"a}mer}},
  \bibinfo {author} {\bibfnamefont {D.}~\bibnamefont {Reith}},\ and\ \bibinfo
  {author} {\bibfnamefont {H.}~\bibnamefont {Foysi}},\ }\bibfield  {title}
  {\enquote {\bibinfo {title} {{Semi-Lagrangian lattice Boltzmann method for
  compressible flows}},}\ }\href@noop {} {\bibfield  {journal} {\bibinfo
  {journal} {Physical Review E}\ }\textbf {\bibinfo {volume} {101}},\ \bibinfo
  {pages} {053306} (\bibinfo {year} {2020})}\BibitemShut {NoStop}%
\bibitem [{\citenamefont {Frapolli}(2017)}]{frapolli2017entropic}%
  \BibitemOpen
  \bibfield  {author} {\bibinfo {author} {\bibfnamefont {N.}~\bibnamefont
  {Frapolli}},\ }\emph {\bibinfo {title} {{Entropic lattice Boltzmann models
  for thermal and compressible flows}}},\ \href@noop {} {Ph.D. thesis},\
  \bibinfo  {school} {ETH Zurich} (\bibinfo {year} {2017})\BibitemShut
  {NoStop}%
\bibitem [{\citenamefont {Prasianakis}\ and\ \citenamefont
  {Karlin}(2008)}]{prasianakis2008lattice}%
  \BibitemOpen
  \bibfield  {author} {\bibinfo {author} {\bibfnamefont {N.~I.}\ \bibnamefont
  {Prasianakis}}\ and\ \bibinfo {author} {\bibfnamefont {I.~V.}\ \bibnamefont
  {Karlin}},\ }\bibfield  {title} {\enquote {\bibinfo {title} {{Lattice
  Boltzmann method for simulation of compressible flows on standard
  lattices}},}\ }\href@noop {} {\bibfield  {journal} {\bibinfo  {journal}
  {Physical Review E}\ }\textbf {\bibinfo {volume} {78}},\ \bibinfo {pages}
  {016704} (\bibinfo {year} {2008})}\BibitemShut {NoStop}%
\bibitem [{\citenamefont {Prasianakis}\ \emph {et~al.}(2009)\citenamefont
  {Prasianakis}, \citenamefont {Karlin}, \citenamefont {Mantzaras},\ and\
  \citenamefont {Boulouchos}}]{prasianakis2009lattice}%
  \BibitemOpen
  \bibfield  {author} {\bibinfo {author} {\bibfnamefont {N.~I.}\ \bibnamefont
  {Prasianakis}}, \bibinfo {author} {\bibfnamefont {I.~V.}\ \bibnamefont
  {Karlin}}, \bibinfo {author} {\bibfnamefont {J.}~\bibnamefont {Mantzaras}},\
  and\ \bibinfo {author} {\bibfnamefont {K.~B.}\ \bibnamefont {Boulouchos}},\
  }\bibfield  {title} {\enquote {\bibinfo {title} {{Lattice Boltzmann method
  with restored Galilean invariance}},}\ }\href@noop {} {\bibfield  {journal}
  {\bibinfo  {journal} {Physical Review E}\ }\textbf {\bibinfo {volume} {79}},\
  \bibinfo {pages} {066702} (\bibinfo {year} {2009})}\BibitemShut {NoStop}%
\bibitem [{\citenamefont {Guo}\ \emph {et~al.}(2007)\citenamefont {Guo},
  \citenamefont {Zheng}, \citenamefont {Shi},\ and\ \citenamefont
  {Zhao}}]{guo2007thermal}%
  \BibitemOpen
  \bibfield  {author} {\bibinfo {author} {\bibfnamefont {Z.}~\bibnamefont
  {Guo}}, \bibinfo {author} {\bibfnamefont {C.}~\bibnamefont {Zheng}}, \bibinfo
  {author} {\bibfnamefont {B.}~\bibnamefont {Shi}},\ and\ \bibinfo {author}
  {\bibfnamefont {T.}~\bibnamefont {Zhao}},\ }\bibfield  {title} {\enquote
  {\bibinfo {title} {{Thermal lattice Boltzmann equation for low Mach number
  flows: decoupling model}},}\ }\href@noop {} {\bibfield  {journal} {\bibinfo
  {journal} {Physical Review E}\ }\textbf {\bibinfo {volume} {75}},\ \bibinfo
  {pages} {036704} (\bibinfo {year} {2007})}\BibitemShut {NoStop}%
\bibitem [{\citenamefont {Feng}, \citenamefont {Sagaut},\ and\ \citenamefont
  {Tao}(2015)}]{feng2015three}%
  \BibitemOpen
  \bibfield  {author} {\bibinfo {author} {\bibfnamefont {Y.}~\bibnamefont
  {Feng}}, \bibinfo {author} {\bibfnamefont {P.}~\bibnamefont {Sagaut}},\ and\
  \bibinfo {author} {\bibfnamefont {W.}~\bibnamefont {Tao}},\ }\bibfield
  {title} {\enquote {\bibinfo {title} {{A three dimensional lattice model for
  thermal compressible flow on standard lattices}},}\ }\href@noop {} {\bibfield
   {journal} {\bibinfo  {journal} {Journal of Computational Physics}\ }\textbf
  {\bibinfo {volume} {303}},\ \bibinfo {pages} {514--529} (\bibinfo {year}
  {2015})}\BibitemShut {NoStop}%
\bibitem [{\citenamefont {Saadat}, \citenamefont {B{\"o}sch},\ and\
  \citenamefont {Karlin}(2019)}]{saadat2019lattice}%
  \BibitemOpen
  \bibfield  {author} {\bibinfo {author} {\bibfnamefont {M.~H.}\ \bibnamefont
  {Saadat}}, \bibinfo {author} {\bibfnamefont {F.}~\bibnamefont {B{\"o}sch}},\
  and\ \bibinfo {author} {\bibfnamefont {I.~V.}\ \bibnamefont {Karlin}},\
  }\bibfield  {title} {\enquote {\bibinfo {title} {{Lattice Boltzmann model for
  compressible flows on standard lattices: Variable Prandtl number and
  adiabatic exponent}},}\ }\href@noop {} {\bibfield  {journal} {\bibinfo
  {journal} {Physical Review E}\ }\textbf {\bibinfo {volume} {99}},\ \bibinfo
  {pages} {013306} (\bibinfo {year} {2019})}\BibitemShut {NoStop}%
\bibitem [{\citenamefont {Hosseini}, \citenamefont {Darabiha},\ and\
  \citenamefont {Th{\'e}venin}(2020)}]{hosseini2020compressibility}%
  \BibitemOpen
  \bibfield  {author} {\bibinfo {author} {\bibfnamefont {S.~A.}\ \bibnamefont
  {Hosseini}}, \bibinfo {author} {\bibfnamefont {N.}~\bibnamefont {Darabiha}},\
  and\ \bibinfo {author} {\bibfnamefont {D.}~\bibnamefont {Th{\'e}venin}},\
  }\bibfield  {title} {\enquote {\bibinfo {title} {{Compressibility in lattice
  Boltzmann on standard stencils: effects of deviation from reference
  temperature}},}\ }\href@noop {} {\bibfield  {journal} {\bibinfo  {journal}
  {Philosophical Transactions of the Royal Society A}\ }\textbf {\bibinfo
  {volume} {378}},\ \bibinfo {pages} {20190399} (\bibinfo {year}
  {2020})}\BibitemShut {NoStop}%
\bibitem [{\citenamefont {Feng}, \citenamefont {Sagaut},\ and\ \citenamefont
  {Tao}(2016)}]{feng2016compressible}%
  \BibitemOpen
  \bibfield  {author} {\bibinfo {author} {\bibfnamefont {Y.}~\bibnamefont
  {Feng}}, \bibinfo {author} {\bibfnamefont {P.}~\bibnamefont {Sagaut}},\ and\
  \bibinfo {author} {\bibfnamefont {W.-Q.}\ \bibnamefont {Tao}},\ }\bibfield
  {title} {\enquote {\bibinfo {title} {{A compressible lattice Boltzmann finite
  volume model for high subsonic and transonic flows on regular lattices}},}\
  }\href@noop {} {\bibfield  {journal} {\bibinfo  {journal} {Computers \&
  Fluids}\ }\textbf {\bibinfo {volume} {131}},\ \bibinfo {pages} {45--55}
  (\bibinfo {year} {2016})}\BibitemShut {NoStop}%
\bibitem [{\citenamefont {Feng}\ \emph {et~al.}(2019)\citenamefont {Feng},
  \citenamefont {Boivin}, \citenamefont {Jacob},\ and\ \citenamefont
  {Sagaut}}]{feng2019hybrid}%
  \BibitemOpen
  \bibfield  {author} {\bibinfo {author} {\bibfnamefont {Y.}~\bibnamefont
  {Feng}}, \bibinfo {author} {\bibfnamefont {P.}~\bibnamefont {Boivin}},
  \bibinfo {author} {\bibfnamefont {J.}~\bibnamefont {Jacob}},\ and\ \bibinfo
  {author} {\bibfnamefont {P.}~\bibnamefont {Sagaut}},\ }\bibfield  {title}
  {\enquote {\bibinfo {title} {{Hybrid recursive regularized thermal lattice
  Boltzmann model for high subsonic compressible flows}},}\ }\href@noop {}
  {\bibfield  {journal} {\bibinfo  {journal} {Journal of Computational
  Physics}\ }\textbf {\bibinfo {volume} {394}},\ \bibinfo {pages} {82--99}
  (\bibinfo {year} {2019})}\BibitemShut {NoStop}%
\bibitem [{\citenamefont {Guo}, \citenamefont {Feng},\ and\ \citenamefont
  {Sagaut}(2020)}]{guo2020improved}%
  \BibitemOpen
  \bibfield  {author} {\bibinfo {author} {\bibfnamefont {S.}~\bibnamefont
  {Guo}}, \bibinfo {author} {\bibfnamefont {Y.}~\bibnamefont {Feng}},\ and\
  \bibinfo {author} {\bibfnamefont {P.}~\bibnamefont {Sagaut}},\ }\bibfield
  {title} {\enquote {\bibinfo {title} {{Improved standard thermal lattice
  Boltzmann model with hybrid recursive regularization for compressible laminar
  and turbulent flows}},}\ }\href@noop {} {\bibfield  {journal} {\bibinfo
  {journal} {Physics of Fluids}\ }\textbf {\bibinfo {volume} {32}},\ \bibinfo
  {pages} {126108} (\bibinfo {year} {2020})}\BibitemShut {NoStop}%
\bibitem [{\citenamefont {Zhao}\ \emph {et~al.}(2020)\citenamefont {Zhao},
  \citenamefont {Farag}, \citenamefont {Boivin},\ and\ \citenamefont
  {Sagaut}}]{zhao2020toward}%
  \BibitemOpen
  \bibfield  {author} {\bibinfo {author} {\bibfnamefont {S.}~\bibnamefont
  {Zhao}}, \bibinfo {author} {\bibfnamefont {G.}~\bibnamefont {Farag}},
  \bibinfo {author} {\bibfnamefont {P.}~\bibnamefont {Boivin}},\ and\ \bibinfo
  {author} {\bibfnamefont {P.}~\bibnamefont {Sagaut}},\ }\bibfield  {title}
  {\enquote {\bibinfo {title} {{Toward fully conservative hybrid lattice
  Boltzmann methods for compressible flows}},}\ }\href@noop {} {\bibfield
  {journal} {\bibinfo  {journal} {Physics of Fluids}\ }\textbf {\bibinfo
  {volume} {32}},\ \bibinfo {pages} {126118} (\bibinfo {year}
  {2020})}\BibitemShut {NoStop}%
\bibitem [{\citenamefont {Li}\ \emph {et~al.}(2012)\citenamefont {Li},
  \citenamefont {Luo}, \citenamefont {He}, \citenamefont {Gao},\ and\
  \citenamefont {Tao}}]{li2012coupling}%
  \BibitemOpen
  \bibfield  {author} {\bibinfo {author} {\bibfnamefont {Q.}~\bibnamefont
  {Li}}, \bibinfo {author} {\bibfnamefont {K.}~\bibnamefont {Luo}}, \bibinfo
  {author} {\bibfnamefont {Y.}~\bibnamefont {He}}, \bibinfo {author}
  {\bibfnamefont {Y.}~\bibnamefont {Gao}},\ and\ \bibinfo {author}
  {\bibfnamefont {W.}~\bibnamefont {Tao}},\ }\bibfield  {title} {\enquote
  {\bibinfo {title} {{Coupling lattice Boltzmann model for simulation of
  thermal flows on standard lattices}},}\ }\href@noop {} {\bibfield  {journal}
  {\bibinfo  {journal} {Physical Review E}\ }\textbf {\bibinfo {volume} {85}},\
  \bibinfo {pages} {016710} (\bibinfo {year} {2012})}\BibitemShut {NoStop}%
\bibitem [{\citenamefont {Karlin}, \citenamefont {Sichau},\ and\ \citenamefont
  {Chikatamarla}(2013)}]{karlin2013consistent}%
  \BibitemOpen
  \bibfield  {author} {\bibinfo {author} {\bibfnamefont {I.}~\bibnamefont
  {Karlin}}, \bibinfo {author} {\bibfnamefont {D.}~\bibnamefont {Sichau}},\
  and\ \bibinfo {author} {\bibfnamefont {S.}~\bibnamefont {Chikatamarla}},\
  }\bibfield  {title} {\enquote {\bibinfo {title} {{Consistent two-population
  lattice Boltzmann model for thermal flows}},}\ }\href@noop {} {\bibfield
  {journal} {\bibinfo  {journal} {Physical Review E}\ }\textbf {\bibinfo
  {volume} {88}},\ \bibinfo {pages} {063310} (\bibinfo {year}
  {2013})}\BibitemShut {NoStop}%
\bibitem [{\citenamefont {Dorschner}, \citenamefont {B{\"o}sch},\ and\
  \citenamefont {Karlin}(2018)}]{dorschner2018particles}%
  \BibitemOpen
  \bibfield  {author} {\bibinfo {author} {\bibfnamefont {B.}~\bibnamefont
  {Dorschner}}, \bibinfo {author} {\bibfnamefont {F.}~\bibnamefont
  {B{\"o}sch}},\ and\ \bibinfo {author} {\bibfnamefont {I.~V.}\ \bibnamefont
  {Karlin}},\ }\bibfield  {title} {\enquote {\bibinfo {title} {Particles on
  demand for kinetic theory},}\ }\href@noop {} {\bibfield  {journal} {\bibinfo
  {journal} {Physical review letters}\ }\textbf {\bibinfo {volume} {121}},\
  \bibinfo {pages} {130602} (\bibinfo {year} {2018})}\BibitemShut {NoStop}%
\bibitem [{\citenamefont {Saadat}, \citenamefont {B{\"o}sch},\ and\
  \citenamefont {Karlin}(2020)}]{saadat2020semi}%
  \BibitemOpen
  \bibfield  {author} {\bibinfo {author} {\bibfnamefont {M.~H.}\ \bibnamefont
  {Saadat}}, \bibinfo {author} {\bibfnamefont {F.}~\bibnamefont {B{\"o}sch}},\
  and\ \bibinfo {author} {\bibfnamefont {I.~V.}\ \bibnamefont {Karlin}},\
  }\bibfield  {title} {\enquote {\bibinfo {title} {{Semi-Lagrangian lattice
  Boltzmann model for compressible flows on unstructured meshes}},}\
  }\href@noop {} {\bibfield  {journal} {\bibinfo  {journal} {Physical Review
  E}\ }\textbf {\bibinfo {volume} {101}},\ \bibinfo {pages} {023311} (\bibinfo
  {year} {2020})}\BibitemShut {NoStop}%
\bibitem [{\citenamefont {Frapolli}, \citenamefont {Chikatamarla},\ and\
  \citenamefont {Karlin}(2016{\natexlab{b}})}]{frapolli2016lattice}%
  \BibitemOpen
  \bibfield  {author} {\bibinfo {author} {\bibfnamefont {N.}~\bibnamefont
  {Frapolli}}, \bibinfo {author} {\bibfnamefont {S.~S.}\ \bibnamefont
  {Chikatamarla}},\ and\ \bibinfo {author} {\bibfnamefont {I.~V.}\ \bibnamefont
  {Karlin}},\ }\bibfield  {title} {\enquote {\bibinfo {title} {Lattice kinetic
  theory in a comoving galilean reference frame},}\ }\href@noop {} {\bibfield
  {journal} {\bibinfo  {journal} {Physical review letters}\ }\textbf {\bibinfo
  {volume} {117}},\ \bibinfo {pages} {010604} (\bibinfo {year}
  {2016}{\natexlab{b}})}\BibitemShut {NoStop}%
\bibitem [{\citenamefont {Bhatnagar}, \citenamefont {Gross},\ and\
  \citenamefont {Krook}(1954)}]{bhatnagar1954model}%
  \BibitemOpen
  \bibfield  {author} {\bibinfo {author} {\bibfnamefont {P.~L.}\ \bibnamefont
  {Bhatnagar}}, \bibinfo {author} {\bibfnamefont {E.~P.}\ \bibnamefont
  {Gross}},\ and\ \bibinfo {author} {\bibfnamefont {M.}~\bibnamefont {Krook}},\
  }\bibfield  {title} {\enquote {\bibinfo {title} {{A model for collision
  processes in gases. I. Small amplitude processes in charged and neutral
  one-component systems}},}\ }\href@noop {} {\bibfield  {journal} {\bibinfo
  {journal} {Physical Review}\ }\textbf {\bibinfo {volume} {94}},\ \bibinfo
  {pages} {511} (\bibinfo {year} {1954})}\BibitemShut {NoStop}%
\bibitem [{\citenamefont {Karlin}\ and\ \citenamefont
  {Asinari}(2010)}]{karlin2010factorization}%
  \BibitemOpen
  \bibfield  {author} {\bibinfo {author} {\bibfnamefont {I.}~\bibnamefont
  {Karlin}}\ and\ \bibinfo {author} {\bibfnamefont {P.}~\bibnamefont
  {Asinari}},\ }\bibfield  {title} {\enquote {\bibinfo {title} {{Factorization
  symmetry in the lattice Boltzmann method}},}\ }\href@noop {} {\bibfield
  {journal} {\bibinfo  {journal} {Physica A: Statistical Mechanics and its
  Applications}\ }\textbf {\bibinfo {volume} {389}},\ \bibinfo {pages}
  {1530--1548} (\bibinfo {year} {2010})}\BibitemShut {NoStop}%
\bibitem [{\citenamefont {Saadat}, \citenamefont {Dorschner},\ and\
  \citenamefont {Karlin}(2021)}]{saadat2021extended}%
  \BibitemOpen
  \bibfield  {author} {\bibinfo {author} {\bibfnamefont {M.~H.}\ \bibnamefont
  {Saadat}}, \bibinfo {author} {\bibfnamefont {B.}~\bibnamefont {Dorschner}},\
  and\ \bibinfo {author} {\bibfnamefont {I.~V.}\ \bibnamefont {Karlin}},\
  }\href@noop {} {\enquote {\bibinfo {title} {{Extended Lattice Boltzmann
  Model}},}\ } (\bibinfo {year} {2021}),\ \Eprint
  {https://arxiv.org/abs/2101.04550} {arXiv:2101.04550 [physics.flu-dyn]}
  \BibitemShut {NoStop}%
\bibitem [{\citenamefont {Sod}(1978)}]{sod1978survey}%
  \BibitemOpen
  \bibfield  {author} {\bibinfo {author} {\bibfnamefont {G.~A.}\ \bibnamefont
  {Sod}},\ }\bibfield  {title} {\enquote {\bibinfo {title} {A survey of several
  finite difference methods for systems of nonlinear hyperbolic conservation
  laws},}\ }\href@noop {} {\bibfield  {journal} {\bibinfo  {journal} {Journal
  of Computational Physics}\ }\textbf {\bibinfo {volume} {27}},\ \bibinfo
  {pages} {1--31} (\bibinfo {year} {1978})}\BibitemShut {NoStop}%
\bibitem [{\citenamefont {Inoue}\ and\ \citenamefont
  {Hattori}(1999)}]{inoue1999sound}%
  \BibitemOpen
  \bibfield  {author} {\bibinfo {author} {\bibfnamefont {O.}~\bibnamefont
  {Inoue}}\ and\ \bibinfo {author} {\bibfnamefont {Y.}~\bibnamefont
  {Hattori}},\ }\bibfield  {title} {\enquote {\bibinfo {title} {Sound
  generation by shock-vortex interactions},}\ }\href@noop {} {\bibfield
  {journal} {\bibinfo  {journal} {Journal of Fluid Mechanics}\ }\textbf
  {\bibinfo {volume} {380}},\ \bibinfo {pages} {81--116} (\bibinfo {year}
  {1999})}\BibitemShut {NoStop}%
\bibitem [{\citenamefont {Lee}, \citenamefont {Lele},\ and\ \citenamefont
  {Moin}(1991)}]{lee1991eddy}%
  \BibitemOpen
  \bibfield  {author} {\bibinfo {author} {\bibfnamefont {S.}~\bibnamefont
  {Lee}}, \bibinfo {author} {\bibfnamefont {S.~K.}\ \bibnamefont {Lele}},\ and\
  \bibinfo {author} {\bibfnamefont {P.}~\bibnamefont {Moin}},\ }\bibfield
  {title} {\enquote {\bibinfo {title} {Eddy shocklets in decaying compressible
  turbulence},}\ }\href@noop {} {\bibfield  {journal} {\bibinfo  {journal}
  {Physics of Fluids A: Fluid Dynamics}\ }\textbf {\bibinfo {volume} {3}},\
  \bibinfo {pages} {657--664} (\bibinfo {year} {1991})}\BibitemShut {NoStop}%
\bibitem [{\citenamefont {Mansour}\ and\ \citenamefont
  {Wray}(1994)}]{mansour1994decay}%
  \BibitemOpen
  \bibfield  {author} {\bibinfo {author} {\bibfnamefont {N.}~\bibnamefont
  {Mansour}}\ and\ \bibinfo {author} {\bibfnamefont {A.}~\bibnamefont {Wray}},\
  }\bibfield  {title} {\enquote {\bibinfo {title} {{Decay of isotropic
  turbulence at low Reynolds number}},}\ }\href@noop {} {\bibfield  {journal}
  {\bibinfo  {journal} {Physics of Fluids}\ }\textbf {\bibinfo {volume} {6}},\
  \bibinfo {pages} {808--814} (\bibinfo {year} {1994})}\BibitemShut {NoStop}%
\bibitem [{\citenamefont {Samtaney}, \citenamefont {Pullin},\ and\
  \citenamefont {Kosovi{\'c}}(2001)}]{samtaney2001direct}%
  \BibitemOpen
  \bibfield  {author} {\bibinfo {author} {\bibfnamefont {R.}~\bibnamefont
  {Samtaney}}, \bibinfo {author} {\bibfnamefont {D.~I.}\ \bibnamefont
  {Pullin}},\ and\ \bibinfo {author} {\bibfnamefont {B.}~\bibnamefont
  {Kosovi{\'c}}},\ }\bibfield  {title} {\enquote {\bibinfo {title} {Direct
  numerical simulation of decaying compressible turbulence and shocklet
  statistics},}\ }\href@noop {} {\bibfield  {journal} {\bibinfo  {journal}
  {Physics of Fluids}\ }\textbf {\bibinfo {volume} {13}},\ \bibinfo {pages}
  {1415--1430} (\bibinfo {year} {2001})}\BibitemShut {NoStop}%
\bibitem [{\citenamefont {Johnsen}\ \emph {et~al.}(2010)\citenamefont
  {Johnsen}, \citenamefont {Larsson}, \citenamefont {Bhagatwala}, \citenamefont
  {Cabot}, \citenamefont {Moin}, \citenamefont {Olson}, \citenamefont {Rawat},
  \citenamefont {Shankar}, \citenamefont {Sj{\"o}green}, \citenamefont {Yee}
  \emph {et~al.}}]{johnsen2010assessment}%
  \BibitemOpen
  \bibfield  {author} {\bibinfo {author} {\bibfnamefont {E.}~\bibnamefont
  {Johnsen}}, \bibinfo {author} {\bibfnamefont {J.}~\bibnamefont {Larsson}},
  \bibinfo {author} {\bibfnamefont {A.~V.}\ \bibnamefont {Bhagatwala}},
  \bibinfo {author} {\bibfnamefont {W.~H.}\ \bibnamefont {Cabot}}, \bibinfo
  {author} {\bibfnamefont {P.}~\bibnamefont {Moin}}, \bibinfo {author}
  {\bibfnamefont {B.~J.}\ \bibnamefont {Olson}}, \bibinfo {author}
  {\bibfnamefont {P.~S.}\ \bibnamefont {Rawat}}, \bibinfo {author}
  {\bibfnamefont {S.~K.}\ \bibnamefont {Shankar}}, \bibinfo {author}
  {\bibfnamefont {B.}~\bibnamefont {Sj{\"o}green}}, \bibinfo {author}
  {\bibfnamefont {H.~C.}\ \bibnamefont {Yee}}, \emph {et~al.},\ }\bibfield
  {title} {\enquote {\bibinfo {title} {Assessment of high-resolution methods
  for numerical simulations of compressible turbulence with shock waves},}\
  }\href@noop {} {\bibfield  {journal} {\bibinfo  {journal} {Journal of
  Computational Physics}\ }\textbf {\bibinfo {volume} {229}},\ \bibinfo {pages}
  {1213--1237} (\bibinfo {year} {2010})}\BibitemShut {NoStop}%
\bibitem [{\citenamefont {Kumar}, \citenamefont {Girimaji},\ and\ \citenamefont
  {Kerimo}(2013)}]{kumar2013weno}%
  \BibitemOpen
  \bibfield  {author} {\bibinfo {author} {\bibfnamefont {G.}~\bibnamefont
  {Kumar}}, \bibinfo {author} {\bibfnamefont {S.~S.}\ \bibnamefont
  {Girimaji}},\ and\ \bibinfo {author} {\bibfnamefont {J.}~\bibnamefont
  {Kerimo}},\ }\bibfield  {title} {\enquote {\bibinfo {title} {{WENO-enhanced
  gas-kinetic scheme for direct simulations of compressible transition and
  turbulence}},}\ }\href@noop {} {\bibfield  {journal} {\bibinfo  {journal}
  {Journal of Computational Physics}\ }\textbf {\bibinfo {volume} {234}},\
  \bibinfo {pages} {499--523} (\bibinfo {year} {2013})}\BibitemShut {NoStop}%
\bibitem [{\citenamefont {Cao}, \citenamefont {Pan},\ and\ \citenamefont
  {Xu}(2019)}]{cao2019three}%
  \BibitemOpen
  \bibfield  {author} {\bibinfo {author} {\bibfnamefont {G.}~\bibnamefont
  {Cao}}, \bibinfo {author} {\bibfnamefont {L.}~\bibnamefont {Pan}},\ and\
  \bibinfo {author} {\bibfnamefont {K.}~\bibnamefont {Xu}},\ }\bibfield
  {title} {\enquote {\bibinfo {title} {{Three dimensional high-order
  gas-kinetic scheme for supersonic isotropic turbulence I: criterion for
  direct numerical simulation}},}\ }\href@noop {} {\bibfield  {journal}
  {\bibinfo  {journal} {Computers \& Fluids}\ }\textbf {\bibinfo {volume}
  {192}},\ \bibinfo {pages} {104273} (\bibinfo {year} {2019})}\BibitemShut
  {NoStop}%
\bibitem [{\citenamefont {Chen}\ \emph {et~al.}(2020)\citenamefont {Chen},
  \citenamefont {Wen}, \citenamefont {Wang}, \citenamefont {Guo}, \citenamefont
  {Wang},\ and\ \citenamefont {Chen}}]{chen2020simulation}%
  \BibitemOpen
  \bibfield  {author} {\bibinfo {author} {\bibfnamefont {T.}~\bibnamefont
  {Chen}}, \bibinfo {author} {\bibfnamefont {X.}~\bibnamefont {Wen}}, \bibinfo
  {author} {\bibfnamefont {L.-P.}\ \bibnamefont {Wang}}, \bibinfo {author}
  {\bibfnamefont {Z.}~\bibnamefont {Guo}}, \bibinfo {author} {\bibfnamefont
  {J.}~\bibnamefont {Wang}},\ and\ \bibinfo {author} {\bibfnamefont
  {S.}~\bibnamefont {Chen}},\ }\bibfield  {title} {\enquote {\bibinfo {title}
  {Simulation of three-dimensional compressible decaying isotropic turbulence
  using a redesigned discrete unified gas kinetic scheme},}\ }\href@noop {}
  {\bibfield  {journal} {\bibinfo  {journal} {Physics of Fluids}\ }\textbf
  {\bibinfo {volume} {32}},\ \bibinfo {pages} {125104} (\bibinfo {year}
  {2020})}\BibitemShut {NoStop}%
\bibitem [{\citenamefont {Meyer}\ and\ \citenamefont
  {Eggersdorfer}(2014)}]{meyer2014simulating}%
  \BibitemOpen
  \bibfield  {author} {\bibinfo {author} {\bibfnamefont {D.~W.}\ \bibnamefont
  {Meyer}}\ and\ \bibinfo {author} {\bibfnamefont {M.~L.}\ \bibnamefont
  {Eggersdorfer}},\ }\bibfield  {title} {\enquote {\bibinfo {title}
  {{Simulating particle collisions in homogeneous turbulence with kinematic
  simulation -- A validation study}},}\ }\href@noop {} {\bibfield  {journal}
  {\bibinfo  {journal} {Colloids and Surfaces A: Physicochemical and
  Engineering Aspects}\ }\textbf {\bibinfo {volume} {454}},\ \bibinfo {pages}
  {57--64} (\bibinfo {year} {2014})}\BibitemShut {NoStop}%
\bibitem [{\citenamefont {Fang}\ \emph {et~al.}(2014)\citenamefont {Fang},
  \citenamefont {Yao}, \citenamefont {Li},\ and\ \citenamefont
  {Lu}}]{fang2014investigation}%
  \BibitemOpen
  \bibfield  {author} {\bibinfo {author} {\bibfnamefont {J.}~\bibnamefont
  {Fang}}, \bibinfo {author} {\bibfnamefont {Y.}~\bibnamefont {Yao}}, \bibinfo
  {author} {\bibfnamefont {Z.}~\bibnamefont {Li}},\ and\ \bibinfo {author}
  {\bibfnamefont {L.}~\bibnamefont {Lu}},\ }\bibfield  {title} {\enquote
  {\bibinfo {title} {Investigation of low-dissipation monotonicity-preserving
  scheme for direct numerical simulation of compressible turbulent flows},}\
  }\href@noop {} {\bibfield  {journal} {\bibinfo  {journal} {Computers \&
  Fluids}\ }\textbf {\bibinfo {volume} {104}},\ \bibinfo {pages} {55--72}
  (\bibinfo {year} {2014})}\BibitemShut {NoStop}%
\bibitem [{\citenamefont {Garnier}\ \emph {et~al.}(1999)\citenamefont
  {Garnier}, \citenamefont {Mossi}, \citenamefont {Sagaut}, \citenamefont
  {Comte},\ and\ \citenamefont {Deville}}]{garnier1999use}%
  \BibitemOpen
  \bibfield  {author} {\bibinfo {author} {\bibfnamefont {E.}~\bibnamefont
  {Garnier}}, \bibinfo {author} {\bibfnamefont {M.}~\bibnamefont {Mossi}},
  \bibinfo {author} {\bibfnamefont {P.}~\bibnamefont {Sagaut}}, \bibinfo
  {author} {\bibfnamefont {P.}~\bibnamefont {Comte}},\ and\ \bibinfo {author}
  {\bibfnamefont {M.}~\bibnamefont {Deville}},\ }\bibfield  {title} {\enquote
  {\bibinfo {title} {On the use of shock-capturing schemes for large-eddy
  simulation},}\ }\href@noop {} {\bibfield  {journal} {\bibinfo  {journal}
  {Journal of Computational Physics}\ }\textbf {\bibinfo {volume} {153}},\
  \bibinfo {pages} {273--311} (\bibinfo {year} {1999})}\BibitemShut {NoStop}%
\bibitem [{\citenamefont {Vuorinen}\ \emph {et~al.}(2013)\citenamefont
  {Vuorinen}, \citenamefont {Yu}, \citenamefont {Tirunagari}, \citenamefont
  {Kaario}, \citenamefont {Larmi}, \citenamefont {Duwig},\ and\ \citenamefont
  {Boersma}}]{vuorinen2013large}%
  \BibitemOpen
  \bibfield  {author} {\bibinfo {author} {\bibfnamefont {V.}~\bibnamefont
  {Vuorinen}}, \bibinfo {author} {\bibfnamefont {J.}~\bibnamefont {Yu}},
  \bibinfo {author} {\bibfnamefont {S.}~\bibnamefont {Tirunagari}}, \bibinfo
  {author} {\bibfnamefont {O.}~\bibnamefont {Kaario}}, \bibinfo {author}
  {\bibfnamefont {M.}~\bibnamefont {Larmi}}, \bibinfo {author} {\bibfnamefont
  {C.}~\bibnamefont {Duwig}},\ and\ \bibinfo {author} {\bibfnamefont
  {B.}~\bibnamefont {Boersma}},\ }\bibfield  {title} {\enquote {\bibinfo
  {title} {Large-eddy simulation of highly underexpanded transient gas jets},}\
  }\href@noop {} {\bibfield  {journal} {\bibinfo  {journal} {Physics of
  Fluids}\ }\textbf {\bibinfo {volume} {25}},\ \bibinfo {pages} {016101}
  (\bibinfo {year} {2013})}\BibitemShut {NoStop}%
\bibitem [{\citenamefont {Pirozzoli}, \citenamefont {Bernardini},\ and\
  \citenamefont {Grasso}(2010)}]{pirozzoli2010direct}%
  \BibitemOpen
  \bibfield  {author} {\bibinfo {author} {\bibfnamefont {S.}~\bibnamefont
  {Pirozzoli}}, \bibinfo {author} {\bibfnamefont {M.}~\bibnamefont
  {Bernardini}},\ and\ \bibinfo {author} {\bibfnamefont {F.}~\bibnamefont
  {Grasso}},\ }\bibfield  {title} {\enquote {\bibinfo {title} {Direct numerical
  simulation of transonic shock/boundary layer interaction under conditions of
  incipient separation},}\ }\href@noop {} {\bibfield  {journal} {\bibinfo
  {journal} {Journal of Fluid Mechanics}\ }\textbf {\bibinfo {volume} {657}},\
  \bibinfo {pages} {361} (\bibinfo {year} {2010})}\BibitemShut {NoStop}%
\end{thebibliography}%

\end{document}